\documentclass{article}
\usepackage{graphicx} % Required for inserting images
\usepackage{lineno,hyperref}
\modulolinenumbers[5]

\usepackage[a4paper, total={7in, 9.6in}]{geometry}
\usepackage{amsmath}
\usepackage{lipsum}
\usepackage{xcolor}
\usepackage{color, colortbl}
\usepackage[framemethod=tikz]{mdframed}
\usepackage{mathtools}
\usepackage{cuted}
\usepackage{float}
\usepackage{ upgreek }
\usepackage{setspace}
\usepackage{amssymb}
\usepackage{float}
\usepackage[utf8]{inputenc}
\usepackage{graphicx}
\usepackage{epstopdf}
\usepackage{tikz}
\usepackage{MnSymbol}
\graphicspath{ {images/} }
\usepackage{cite}
\usepackage{hyperref}
\usepackage[thin, , thinc]{esdiff}
\usepackage{subcaption}
\usepackage{caption}
\usepackage{framed} 
\usepackage{longtable}
\usepackage{url}

\usepackage{alphalph}
\usepackage{booktabs}
\usepackage{dblfloatfix}
\usepackage{placeins}
\renewcommand{\arraystretch}{1.5}
\newcommand{\RN}[1]{ %
	\textup{\uppercase\expandafter{\romannumeral#1}}%
}
\usepackage[labelformat=simple]{subcaption}

\usepackage{amsthm}
\usepackage{verbatim}
\usepackage{color}
\usepackage[ruled,vlined]{algorithm2e}
\usepackage{xcolor}
\usepackage{soul}
\title{Lagrangian evaluation of polymeric stress in viscoelastic fluids
}

\author{
Mohammad Majidi$^{1}$,
Rishu Gandhi$^{2}$,\\
Louison Thorens$^{3,4}$,
Maliheh Teimouri$^{3}$,
Jeffrey S. Guasto$^{3}$,
Arezoo M. Ardekani$^{1*}$
}

\date{
$^{1}$Department of Mechanical Engineering, Purdue University, West Lafayette, IN 47907\\
$^{2}$Department of Mathematics, Applied Science Cluster, School of Advanced Engineering, Dehradun, Uttarakhand, India, 248007\\
$^{3}$Department of Mechanical Engineering, Tufts University, Medford, MA 02155 \\
$^{4}$Institute for Mechanobiology, Department of Bioengineering, College of Engineering, Northeastern University; Boston, MA 02115 
$^*$ Corresponding author: ardekani@purdue.edu}

\begin{document}

\maketitle

% word limit = 250 for JFM
\begin{abstract}

Polymeric stresses in viscoelastic flows arise from the deformation of polymer chains and are commonly computed using Eulerian constitutive models, in which the conformation tensor is evolved as a transported field over the entire domain. This approach is computationally intensive, prone to numerical instabilities, and not directly applicable to experimentally measured velocity fields. In this work,  we develop a Lagrangian integration scheme that reconstructs the polymeric stress field from the deformation-gradient history along fluid element trajectories in a known, steady velocity field. This approach avoids solving the full Eulerian constitutive transport equation, which we develop for the nonlinear FENE-P model as well as the Oldroyd-B model as a reference case. 
After validation on unidirectional, canonical flows, the scheme is applied to non-trivial channel flows past circular obstacles using velocity fields quantified from both numerical simulations and microfluidic experiments.
The reconstructed stress fields across both experiments and simulations are in agreement with traditional Eulerian reference solutions. Not only does this new Lagrangian scheme enable the quantification of stress fields directly from experimental velocity field data, but it also enables partial or whole-field mapping of stresses without solving fully-coupled viscoelastic constitutive equations.

\end{abstract}

\section{Introduction}

Viscoelastic flows impact wide-ranging biological, industrial, and geophysical applications, including enhanced oil recovery, microfluidic mixing, manufacturing, and blood flow \cite{chen2010rheology, sorbie2013polymer, groisman2001efficient, thiebaud2014prediction, kumar2022transport}.
The deformation and relaxation of polymer chains play a critical role in determining flow resistance, transport, mixing, and the onset of elastic flow instabilities in these systems \cite{groisman2000elastic, virk1975drag, white2008mechanics, larson1992instabilities}.  In viscoelastic flows, the polymeric stresses depend not only on the rate of fluid deformation, but also on the deformation history experienced by the polymer chains as they are advected with the flow. The relative importance  of elastic and viscous effects in these flows is commonly characterized by the Weissenberg number, \(\mathrm{Wi}=\lambda_0 \dot{\gamma}\), where \(\lambda_0\) and \(\dot{\gamma}\) denote the polymer relaxation time and a characteristic deformation rate, respectively  \cite{kumarpof, kumar2022transport}. 
 
Unlike Newtonian fluids, the history dependence of polymeric stresses has motivated the development of constitutive models that evolve the fluid microstructure under advection, stretching, and relaxation \cite{bird1987dynamics,ottinger2012stochastic,larson2013constitutive}. 
A  major challenge in quantifying viscoelastic flows is that the evolution of the conformation tensor is commonly evaluated in an Eulerian form using upper-convected derivatives. 
This formulation makes the accumulated deformation history of the fluid elements non-intuitive and more challenging to interpret. These derivatives account for the advection, rotation, and stretching of the polymer microstructure in the moving flow, 
but they do not explicitly capture the path-dependent deformation history of individual fluid elements. Numerically, the evolution of the conformation tensor is computed by solving the coupled flow and constitutive equations in an Eulerian framework using finite element or finite volume methods. However, when the velocity field is known, the stress calculation can also be viewed in a physically-intuitive Lagrangian form by following the deformation history of fluid elements, connecting Eulerian and Lagrangian descriptions of viscoelastic materials \cite{snoeijer2020relationship}.

Recent work by Kumar et al. \cite{kumar2023lagrangian, kumar_stress_2023} directly demonstrated the important connection between flow kinematics and stress field structure by showing that the topology of the polymeric stress field closely follows the Lagrangian stretching field, \(S\). The stretching field is computed directly from velocity data through the deformation-gradient tensor and has a well-developed foundation in the literature on Lagrangian coherent structures and finite-time stretching fields \cite{HALLER2001248,voth2002experimental, haller2023transport}. These results demonstrate that Lagrangian analysis can provide valuable insight into viscoelastic stress distributions even when direct stress measurements are unavailable. However, the resulting description is primarily topological in nature: it reveals the structure of the stress field, but does not directly provide the constitutive stress field itself.

Akin to Lagrangian stretching, Stone et al. \cite{stone2023note} presented a direct material-line derivation of the convected derivatives commonly used in complex-fluid models, providing physically-intuitive insight into microstructure deformation in viscoelastic fluids. Furthermore, by integrating a constitutive model, this approach illustrates the utility of the Lagrangian approach in analyzing viscoelastic stress formation, whereby polymeric stresses depend on their deformation history.   Their Lagrangian integration was developed for the Oldroyd-B model, one of the simplest viscoelastic constitutive models, but it is widely recognized that more sophisticated models are often needed to capture important features of polymeric stress, for example finite polymer extensibility represented in the FENE-P model \cite{boyko2024perspective}. 

Direct experimental measurements of polymeric stress fields remains challenging, emphasizing the need for new computational and hybrid approaches. Rheo-optical methods such as flow-induced birefringence can provide spatially resolved information about stress or molecular alignment \cite{Fuller_1995,Murphy_Davidson_2013}, but they require specialized instrumentation, rely on material-dependent optical response, and may become difficult to interpret when the stress-optical relation breaks down \cite{ober2011spatially, sun2016measurements}. Direct visualization of individual polymer molecules reveal only local conformation dynamics in ideal flows \cite{Perkins1997, Smith1999} and porous geometries \cite{kawale2017polymer}, but such measurements are experimentally demanding and do not directly provide a practical route for whole-field constitutive stress evaluation. By contrast, velocity fields are far more accessible experimentally through techniques such as particle image velocimetry (PIV) \cite{Devasenathipathy2002}.
Thus, the potential to evaluate viscoelastic stresses from measured velocity fields is highly appealing. In the Lagrangian form, the stress can be calculated along selected trajectories from the deformation history of the fluid elements. 

 Motivated by these previous studies, the goal of the present work is to extend the Lagrangian stress-reconstruction framework toward direct evaluation of the conformation tensor and polymeric stress for the nonlinear FENE-P model using the deformation history of fluid particles. The Oldroyd-B model is also included for comparsion as a simpler reference case for the FENE-P results. This approach builds on the idea of Lagranian fluid stretching and the kinematic interpretation of convected derivatives \cite{stone2023note}. First, the method is studied for the case of steady shear flow and compared to existing analytical and asymptotic solutions \cite{sureshkumar1995linear, yamani2023master}. Next, we extend the approach to quantify stress fields in planar channel flow for both rheological models. Finally, we examine channel flows with one and two cylindrical obstacles. For the latter channel flows, velocity and Lagrangian stretching fields obtained from simulations and microfluidic experiments are compared to illustrate flow kinematics. Lagrangian stress fields are reconstructed from both simulations and experiments, and compared with Eulerian simulation results for reference. In this way, the present study provides a direct link between experimentally accessible flow kinematics and viscoelastic stress development in complex flows.

\section{Mathematical Formulation}

 The Lagrangian formulation developed in this work is used to compute the conformation tensor and polymeric stress as presented below. Starting from the Eulerian constitutive equation, a material element is followed through the flow in order to switch from an Eulerian to a Lagrangian description. The formulation is written in terms of the deformation gradient tensor. The resulting expression for the conformation tensor is exact but implicit for the FENE-P model, because it depends on its own trace. Therefore, for the FENE-P model, the trace of the conformation tensor is determined iteratively, after which the full conformation tensor is reconstructed and the polymeric stress is computed from the constitutive stress equation.

\subsection{Lagrangian Integration of the Conformation Tensor and Polymeric Stress}

The conformation tensor equation  written in the general form \cite{bird1987dynamics} is:
\begin{equation}\label{111}
    \mathbf{A}_{(1)}=-\frac{1}{\lambda_0}[f(\operatorname{tr}( \mathbf{A})) \mathbf{A}-a \mathbf{I}],
\end{equation}
where \(\mathbf{A}\) is the conformation tensor, \(\lambda_0\) is the dumbbell relaxation time, and \(\mathbf{I}\) is the identity tensor. The upper-convected derivative of \(\mathbf{A}\) is defined as:
\begin{equation}\label{ucdd}
    \mathbf{A}_{(1)}
    =
    \frac{\partial \mathbf{A}}{\partial t}
    +\mathbf{u}\cdot\nabla \mathbf{A}
    -(\nabla \mathbf{u})^\top\mathbf{A}
    -\mathbf{A}\nabla \mathbf{u}.
\end{equation}
The general form of equation~(\ref{111}) encompasses both the Oldroyd-B model (\(f=1\), \(a=1\)), and the FENE-P model, where the finite extensibility function is given by:
\begin{equation*}
    f(\operatorname{tr}(\mathbf{A}))=\frac{L^2}{L^2-\operatorname{tr}(\mathbf{A})},
\end{equation*}
with \(a=L^2 / (L^2-\operatorname{tr}(\mathbf{I}))\) and \(L\) is the extensibility parameter \cite{bird1980polymer, bird1987dynamics}.  
To recast equation~(\ref{ucdd}) in Lagrangian form, let \(\mathbf{x}(t;\mathbf{X})\) denote the trajectory of a material element initially located at \(\mathbf{X}\), and let \(\boldsymbol{\chi}(\mathbf{X},t)\) denote the associated flow map. The deformation gradient tensor is then defined as:
\begin{equation}
    \mathbf{F}(t;\mathbf{X})=\nabla_{\mathbf{X}} \boldsymbol{\chi}(\mathbf{X},t),
\end{equation}
which maps an infinitesimal material line element in the reference configuration to its deformed state at time, \(t\). Along the same material trajectory, the conformation tensor is written in Lagrangian form as \(\mathbf{A}_L(t;\mathbf{X})=\mathbf{A}(\mathbf{x}(t;\mathbf{X}),t)\).

Using this Lagrangian description, the upper-convected derivative in Lagrangian form is:
\begin{equation}\label{2e}
    {\mathbf{A}_L}_{(1)}
    =
    \frac{\partial \mathbf{A}_L}{\partial t}
    +\mathbf{F}^\top\frac{\partial \mathbf{F}^{-\top}}{\partial t}\mathbf{A}_L
    +\mathbf{A}_L\frac{\partial \mathbf{F}^{-1}}{\partial t}\mathbf{F}.
\end{equation}
Substituting equation~(\ref{2e}) into equation~(\ref{111}) gives:
\begin{equation}\label{55}
    \frac{\partial \mathbf{A}_L}{\partial t}
    +\mathbf{F}^\top\frac{\partial \mathbf{F}^{-\top}}{\partial t}\mathbf{A}_L
    +\mathbf{A}_L\frac{\partial \mathbf{F}^{-1}}{\partial t}\mathbf{F}
    =
    -\frac{1}{\lambda_0}
    \left[
    f(\operatorname{tr}(\mathbf{A}_L))\mathbf{A}_L
    -a\mathbf{I}
    \right].
\end{equation}
Then, equation~(\ref{55}) is left-multiplied by \(\mathbf{F}^{-\top}\) and right-multiplied by \(\mathbf{F}^{-1}\) giving:
\begin{equation*}
     \mathbf{F}^{-\top}\frac{\partial \mathbf{A}_L}{\partial t}\mathbf{F}^{-1}
    +\frac{\partial \mathbf{F}^{-\top}}{\partial t}\mathbf{A}_L\mathbf{F}^{-1}
    +\mathbf{F}^{-\top}\mathbf{A}_L\frac{\partial \mathbf{F}^{-1}}{\partial t} 
    =
    -\frac{1}{\lambda_0}\mathbf{F}^{-\top}
    \left[
    f(\operatorname{tr}(\mathbf{A}_L))\mathbf{A}_L
    -a\mathbf{I}
    \right]\mathbf{F}^{-1}.
\end{equation*}
Applying the product rule to the quantity \(\mathbf{F}^{-\top}\mathbf{A}_L\mathbf{F}^{-1}\), we obtain:
\begin{equation*}
    \frac{\partial}{\partial t}(\mathbf{F}^{-\top}\mathbf{A}_L\mathbf{F}^{-1})
    =
    \frac{\partial \mathbf{F}^{-\top}}{\partial t}\mathbf{A}_L\mathbf{F}^{-1}
    +\mathbf{F}^{-\top}\frac{\partial \mathbf{A}_L}{\partial t}\mathbf{F}^{-1}
    +\mathbf{F}^{-\top}\mathbf{A}_L\frac{\partial \mathbf{F}^{-1}}{\partial t}.
\end{equation*}
Therefore, equation~(\ref{55}) becomes:
\begin{equation*}
    \frac{\partial}{\partial t}(\mathbf{F}^{-\top}\mathbf{A}_L\mathbf{F}^{-1})
    =
    -\frac{1}{\lambda_0}\mathbf{F}^{-\top}
    \left[
    f(\operatorname{tr}(\mathbf{A}_L))\mathbf{A}_L
    -a\mathbf{I}
    \right]\mathbf{F}^{-1},
\end{equation*}
and after simplifying, we obtain:
\begin{equation}\label{666}
    \frac{\partial}{\partial t}
    \left(\mathbf{F}^{-\top}\mathbf{A}_L\mathbf{F}^{-1}\right)
    +
    \frac{1}{\lambda_0}
    f(\operatorname{tr}(\mathbf{A}_L))
    \left(\mathbf{F}^{-\top}\mathbf{A}_L\mathbf{F}^{-1}\right)
    =
    \frac{a}{\lambda_0}
    \mathbf{F}^{-\top}\mathbf{F}^{-1}.
\end{equation}

Equation~(\ref{666}) is a linear first-order tensor-valued ordinary differential equation for 
\(\mathbf{F}^{-\top}\mathbf{A}_L\mathbf{F}^{-1}\). Using an integrating factor, its solution is:
\begin{equation}\label{5e}
\begin{aligned}
    \mathbf{F}^{-\top}(t)\mathbf{A}_L(t)\mathbf{F}^{-1}(t)
    ={}& \exp\!\left(
    -\int_0^t 
    \frac{1}{\lambda_0}f(\operatorname{tr}(\mathbf{A}_L(s)))\,ds
    \right) \\
    &\times
    \bigg[
    \int_{0}^{t}
    \exp\!\left(
    \int_0^{t'} 
    \frac{1}{\lambda_0}f(\operatorname{tr}(\mathbf{A}_L(s)))\,ds
    \right)
    \frac{a}{\lambda_0}
    \mathbf{F}^{-\top}(t')\mathbf{F}^{-1}(t')\,dt'
    +\mathbf{C}
    \bigg],
\end{aligned}
\end{equation}
where \(\mathbf{C}\) is a constant tensor determined by the initial condition.
Taking \(\mathbf{F}(0)=\mathbf{I}\), we have
\[
\mathbf{F}^{-\top}(0)\mathbf{A}_L(0)\mathbf{F}^{-1}(0)=\mathbf{A}_L(0),
\]
and \(\mathbf{C}=\mathbf{A}_L(0)\). Thus,
\begin{equation}\label{6e}
\begin{aligned}
    \mathbf{F}^{-\top}(t)\mathbf{A}_L(t)\mathbf{F}^{-1}(t)
    ={}& \mathbf{A}_L(0)
    \exp\!\left(
    -\int_0^t
    \frac{1}{\lambda_0}f(\operatorname{tr}(\mathbf{A}_L(s)))\,ds
    \right) \\
    &+
    \frac{a}{\lambda_0}
    \exp\!\left(
    -\int_0^t
    \frac{1}{\lambda_0}f(\operatorname{tr}(\mathbf{A}_L(s)))\,ds
    \right)
    \int_{0}^{t}
    \exp\!\left(
    \int_0^{t'}
    \frac{1}{\lambda_0}f(\operatorname{tr}(\mathbf{A}_L(s)))\,ds
    \right)
    \mathbf{F}^{-\top}(t')\mathbf{F}^{-1}(t')\,dt'.
\end{aligned}
\end{equation}

Multiplying from the left by \(\mathbf{F}^\top(t)\) and from the right by \(\mathbf{F}(t)\), we obtain the final representation of the Lagrangian conformation tensor:
% \begin{multline}\label{7e}
%     \mathbf{A}_L(t)
%     =
%     \mathbf{F}^{\top}(t)\mathbf{A}_L(0)\mathbf{F}(t)
%     \exp\!\left(
%     -\int_0^t
%     \frac{1}{\lambda_0}f(\operatorname{tr}(\mathbf{A}_L(s)))\,ds
%     \right) \\
%     +
%     \frac{a}{\lambda_0}
%     \mathbf{F}^{\top}(t)
%     \exp\!\left(
%     -\int_0^t
%     \frac{1}{\lambda_0}f(\operatorname{tr}(\mathbf{A}_L(s)))\,ds
%     \right)
%     \left(
%     \int_{0}^{t}
%     \exp\!\left(
%     \int_0^{t'}
%     \frac{1}{\lambda_0}f(\operatorname{tr}(\mathbf{A}_L(s)))\,ds
%     \right)
%     \mathbf{F}^{-\top}(t')\mathbf{F}^{-1}(t')\,dt'
%     \right)
%     \mathbf{F}(t).
% \end{multline}
 
% For compactness, equation~(\ref{7e}) can be written using the memory factor \(\mathcal{M}(t)\) as:
\begin{equation}\label{7e}
\begin{aligned}
    \mathbf{A}_L(t)
    ={}& \mathbf{F}^{\top}(t)\mathbf{A}_L(0)\mathbf{F}(t)\mathcal{M}(t)  \\
    &+\frac{a}{\lambda_0}
    \mathbf{F}^{\top}(t)\mathcal{M}(t)
    \left(
    \int_{0}^{t}
    \frac{1}{\mathcal{M}(t')}
    \mathbf{F}^{-\top}(t')\mathbf{F}^{-1}(t')\,dt'
    \right)
    \mathbf{F}(t),
\end{aligned}
\end{equation}
where
\[
\mathcal{M}(t)
=
\exp\!\left(
-\int_0^t
\frac{1}{\lambda_0}f(\operatorname{tr}(\mathbf{A}_L(s)))\,ds
\right).
\] 
Equation~(\ref{7e}) provides an exact representation of \(A_L(t)\). For the two-dimensional case, the component-wise form of this expression is given in the Appendix. 
 For the simple case of Oldroyd-B model, where \(f=1\) and \(a=1\), the memory factor reduces to \(\mathcal{M}(t)=e^{-t/\lambda_0}\), and equation~(\ref{7e}) becomes:
\begin{equation}
\begin{aligned}
    \mathbf{A}_L(t)
    ={}& e^{-t/\lambda_0}\mathbf{F}^{\top}(t)\mathbf{A}_L(0)\mathbf{F}(t) \\
    &+\frac{1}{\lambda_0}\mathbf{F}^{\top}(t)
    \left(
    \int_0^t
    \mathbf{F}^{-\top}(t')\mathbf{F}^{-1}(t')
    e^{-(t-t')/\lambda_0}\,dt'
    \right)
    \mathbf{F}(t).
\end{aligned}
\label{eq:oldroydB_lagrangian}
\end{equation}
Therefore, the present formulation recovers the special case of an explicit equation for Oldroyd-B, which was derived in previous work \cite{stone2023note}. The FENE-P case remains implicit, because the memory factor depends on \(\operatorname{tr}(\mathbf{A}_L)\).  

To summarize, the velocity field is first used to reconstruct \(\mathbf{A}_L(t)\) through the deformation-gradient history, and the polymeric stress is then computed from the constitutive stress relation. For the general form considered here, the polymeric stress tensor is given by:
\begin{equation}
    \boldsymbol{\tau}_p
    =
    \frac{\eta_p}{\lambda_0}
    \left[
    f(\operatorname{tr}(\mathbf{A}_L))\mathbf{A}_L
    -
    a\mathbf{I}
    \right],
    \label{eq:stress_relation}
\end{equation}
where \(\eta_p\) is the polymer contribution to the viscosity.

\subsection{Nonlinear Trace Equation and Iterative Solution}
\label{iteratviesolverfene}

In terms of the deformation-gradient history and the scalar quantity \(\operatorname{tr}(\mathbf{A}_L(t))\), equation~(\ref{7e}) gives an exact representation of the Lagrangian conformation tensor \(\mathbf{A}_L(t)\). The remaining difficulty is that the trace appears inside the memory factor through \(f(\operatorname{tr}(\mathbf{A}_L))\). Therefore, although equation~(\ref{7e}) provides the full tensor \(\mathbf{A}_L(t)\), it is not closed. For the two-dimensional case, the diagonal components \(A_{11}(t;T)\) and \(A_{22}(t;T)\) are written explicitly from the component-wise form of equation~(\ref{7e}), given in the Appendix, in terms of the deformation-gradient history and a prescribed trial value \(T\) of the trace. Taking the trace of the tensor representation then leads to a nonlinear scalar equation,
\[
T
=
A_{11}(t;T)+A_{22}(t;T).
\]
%where \(T\) denotes a trial value for \(\operatorname{tr}(\mathbf{A}_L)(t)\). 
The diagonal components on the right-hand side are evaluated from equation~(\ref{7e}) for the same trial value \(T\). To solve this scalar nonlinear problem, we define the residual
\[
R(T)=T-\Big(A_{11}(t;T)+A_{22}(t;T)\Big).
\]
The admissible solution must satisfy
\[
\operatorname{tr}(\mathbf{I})<T<L^2,
\]
which provides a natural interval for the root search. The nonlinear equation \(R(T)=0\) is solved using Brent's method \cite{brent1973algorithms}. Once the converged value \(T^\star=\operatorname{tr}(\mathbf{A}_L)(t)\) is obtained, the conformation tensor \(\mathbf{A}_L(t)\) is reconstructed from equation~(\ref{7e}). Finally, the polymeric stress field is then evaluated from the corresponding stress relation.

\begin{algorithm}[H]
\DontPrintSemicolon

\SetKwBlock{Init}{Initialization}{end}
\Init{
  Specify material parameters and the constitutive model\;
  Provide deformation-gradient history \(\mathbf{F}(t)\) at each material point\;
  Choose admissible bounds for the trace such that \(\operatorname{tr}(\mathbf{I})<T<L^2\)\;
}

\SetKwBlock{TraceSolve}{Trace solution at each material point and time}{end}
\TraceSolve{
  Evaluate \(A_{11}(t;T)\) and \(A_{22}(t;T)\) from the component-wise form of equation~(\ref{7e})\;
  Define the residual function
  \[
  R(T)=T-\Big(A_{11}(t;T)+A_{22}(t;T)\Big)
  \]\;
  Solve \(R(T)=0\) for \(T=\operatorname{tr}(\mathbf{A}_L)(t)\) using Brent's method\;
}

\SetKwBlock{Update}{Tensor and stress reconstruction}{end}
\Update{
  Reconstruct the conformation tensor \(\mathbf{A}_L(t)\) from equation~(\ref{7e}) using the converged trace\;
  Evaluate the polymeric stress from the corresponding stress relation, equation~(\ref{eq:stress_relation})\;
}

\caption{Algorithm for solving the equation for the trace of conformation tensor and reconstructing the conformation tensor and polymeric stress using the Lagrangian approach.}
\label{alg:traceSolve}
\end{algorithm}

\section{Numerical and Experimental Methods}
\label{sec:numerical and exp methods}

\subsection{Numerical Evaluation of the Deformation Gradient Tensor and Stretching Field}

\begin{figure*}[htb]
    \centering
    \includegraphics[width=0.5\textwidth, trim={0cm 0cm 0cm 0cm},clip]{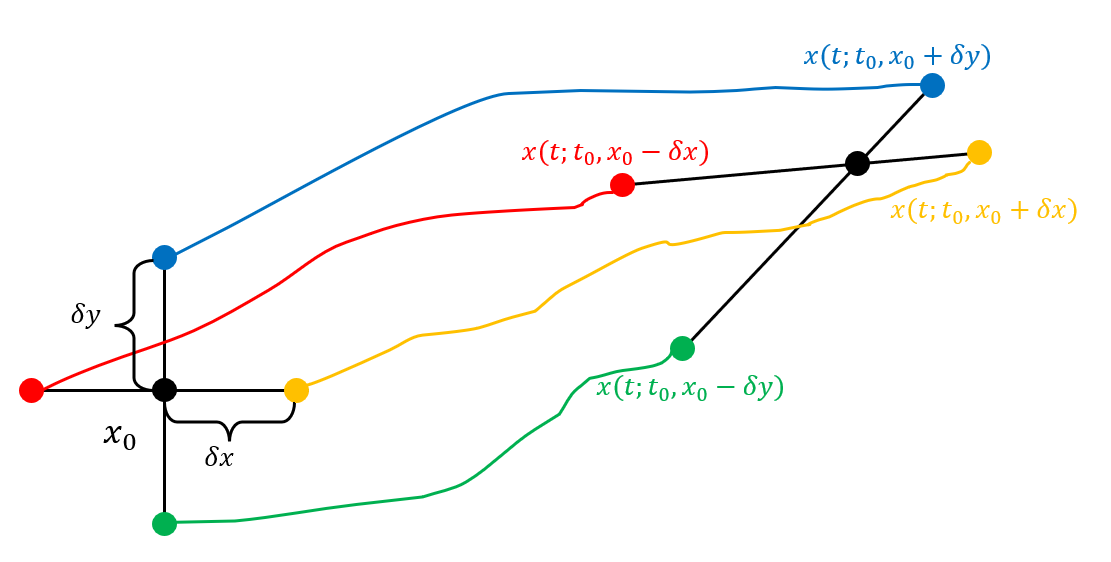}
    \caption{Computation of the deformation gradient tensor at \(\mathbf{x}_0\) using a central point and four auxiliary points initially placed at offsets \(\pm \delta x\) and \(\pm \delta y\). The mapped positions of these auxiliary points at time \(t\) are used to approximate the local deformation gradient by central differencing.}
    \label{fig:Fgraph}
\end{figure*}

 The deformation gradient tensor, \(\mathbf{F}\), which is defined above and provides a mapping between reference frames, is central to the Lagrangian evaluation of the stress field. 
The Lagrangian stretching field, \(S\), is directly related to \(\mathbf{F}\), and this scalar quantity has been shown to mirror the stress field topology. Physically, $S$ measures the maximum accumulated stretch experienced by an infinitesimal material line element along its trajectory over time \cite{kumar_stress_2023,kumar2023lagrangian}. 
Although it does not incorporate fluid rheology, Lagrangian stretching provides 
a valuable kinematic basis for comparing simulated and experimental flow fields by quantifying the deformation experienced by polymers as they are transported through the flow (see Section~\ref{sec:results and discussion}).  \(\mathbf{F}\) and \(S\) are evaluated numerically from the velocity field by considering four auxiliary points placed at small offsets \(\pm \delta x\) and \(\pm \delta y\) around each grid point, \(\mathbf{x}_0\), as illustrated in figure~\ref{fig:Fgraph}. Together with the central grid point, these auxiliary points define a local fluid element.
In the present work, the flow map is obtained numerically by integrating the central and auxiliary points through the velocity field over the selected time interval. The deformation gradient tensor at the primary grid point is then approximated by central differencing of the mapped auxiliary points \cite{Haller2015}. For a two-dimensional flow, this gives
\[
\mathbf{F}(\mathbf{x}_0,t)=
\begin{pmatrix}
\dfrac{x_1(t;t_0,\mathbf{x}_0+\delta x)-x_1(t;t_0,\mathbf{x}_0-\delta x)}{2|\delta x|}
&
\dfrac{x_1(t;t_0,\mathbf{x}_0+\delta y)-x_1(t;t_0,\mathbf{x}_0-\delta y)}{2|\delta y|}
\\[10pt]
\dfrac{x_2(t;t_0,\mathbf{x}_0+\delta x)-x_2(t;t_0,\mathbf{x}_0-\delta x)}{2|\delta x|}
&
\dfrac{x_2(t;t_0,\mathbf{x}_0+\delta y)-x_2(t;t_0,\mathbf{x}_0-\delta y)}{2|\delta y|}
\end{pmatrix}.
\]

Thus, each column of \(\mathbf{F}\) is obtained from the deformation of an initially infinitesimal line element aligned with the corresponding coordinate direction. This procedure provides the deformation history required for the Lagrangian evaluation of the conformation tensor. The stretching field is computed from the right Cauchy--Green deformation tensor,
\[
\mathbf{C_R}(\mathbf{x}_0,t)=\mathbf{F}^\top(\mathbf{x}_0,t)\mathbf{F}(\mathbf{x}_0,t).
\]
The maximum stretching over the selected integration time is then defined as
\[
S(\mathbf{x}_0,t)=\sqrt{\lambda_{\max}\left(\mathbf{C_R}(\mathbf{x}_0,t)\right)},
\]
where \(\lambda_{\max}\) is the largest eigenvalue of \(\mathbf{C_R}\). The deformation gradient tensor and the stretching field reported in the present results for the channel flow with circular obstacles were computed via this approach using the LCS Tool solver \cite{ONU201526}.

\subsection{Eulerian Numerical Solution of the Flow and Polymeric Stress Fields}

 A separate Eulerian simulation is performed to provide the velocity and a reference solution for polymeric stress fields. This solution is used later to assess the stress field reconstructed from the Lagrangian formulation. In the Eulerian simulation, the incompressible flow equations are solved together with the viscoelastic constitutive equation. The flow is governed by the conservation of mass and momentum, 
\begin{equation}
    \nabla \cdot \mathbf{u} = 0,
\end{equation}
\begin{equation}
    \rho
    \left(
    \frac{\partial \mathbf{u}}{\partial t}
    +
    \mathbf{u}\cdot\nabla\mathbf{u}
    \right)
    =
    -\nabla p
    +
    \nabla\cdot\boldsymbol{\tau},
\end{equation}
where \(\mathbf{u}\) is the velocity field, \(p\) is the pressure, \(\rho\) is the fluid density, and \(\boldsymbol{\tau}\) is the total stress tensor. The total stress is written as
\begin{equation}
    \boldsymbol{\tau}
    =
    \boldsymbol{\tau}_s
    +
    \boldsymbol{\tau}_p ,
\end{equation}
where \(\boldsymbol{\tau}_p \) is the polymeric stress tensor, the solvent stress tensor is
\begin{equation}
    \boldsymbol{\tau}_s
    =
    \eta_s
    \left[
    \nabla\mathbf{u}
    +
    \left(\nabla\mathbf{u}\right)^T
    \right],
\end{equation}
and \(\eta_s\) is the solvent viscosity. The numerical simulations are performed in OpenFOAM \cite{jasak2007openfoam} using RheoTool \cite{PIMENTA201785}. In these simulations, the polymeric stress tensor is calculated using the FENE-P constitutive model, which corresponds to a generalized viscoelastic fluid with shear-thinning used in microfluidic experiments (see Section \ref{sec:experimental methods}). The Oldroyd-B model is recovered as the limiting case by taking \(f=1\) and \(a=1\), which provides a good approximation for the Boger fluid used in the experiments \cite{campo-deano_flow_2011}. In stress form, the FENE-P constitutive equation can be written as
\begin{equation}
\boldsymbol{\tau}_p
+
\frac{\lambda_0}{f}
\overset{\nabla}{\boldsymbol{\tau}}_p
=
\frac{a\ \eta_p}{f}
\left[
\nabla\mathbf{u}
+
\left(\nabla\mathbf{u}\right)^\top
\right]
-
\frac{D}{Dt}
\left(
\frac{1}{f}
\right)
\left(
\lambda_0\boldsymbol{\tau}_p
+
\eta_p a\,\mathbf{I}
\right).
\end{equation}
In the numerical implementation, RheoTool uses the log-conformation formulation to improve numerical stability. The solved variable is the logarithm of the conformation tensor,
\begin{equation}
    \boldsymbol{\Theta}
    =
    \log(\mathbf{A}),
\end{equation}
where \(\mathbf{A}\) is the conformation tensor. The polymeric stress tensor is then reconstructed from this variable as
\begin{equation}
    \boldsymbol{\tau}_p
    =
    \frac{\eta_p}{\lambda_0}
    \left[
    f e^{\boldsymbol{\Theta}}
    -
    a\mathbf{I}
    \right].
\end{equation}

\subsection{Experimental Methods}
\label{sec:experimental methods}

For microfluidic experiments, polymer solutions were prepared by dissolving high molecular weight hydrolyzed polyacrylamide (HPAA; $M_w = 18 \times 10^6$; Polysciences, Inc., cat.~no.~18522) at a concentration of 300~ppm ($c/c^* = 0.43$) in a viscous solvent (90\% w/w aqueous glycerol). 
To investigate the purely viscoelastic flows in the absence of shear-thinning, salt (1\% w/w NaCl \cite{browne_bistability_2020}) was added to one HPAA solution to approximate a Boger fluid \cite{campo-deano_flow_2011}. 
%The Newtonian reference fluid was an aqueous glycerol mixtures (90\% w/w glycerol). 
The rheological properties of the solutions were characterized using both shear and extensional rheometry. Shear rheology was performed using a stress-controlled rheometer (HR-20, TA Instruments) with a cone-and-plate geometry to measure viscosity as a function of shear rate under steady conditions. Prior to measurements, polymer solutions were pre-sheared at $1~\mathrm{s^{-1}}$ for 120~s. For each tested shear rate ($0.01\mathrm{~s^{-1}} \le \dot{\gamma} \le 100\mathrm{~s^{-1}}$), the shear rate, $\dot{\gamma}$, was applied for 60~s, and data were subsequently collected for 15~s. The shear viscosity data were parameterized by fitting the Carreau--Yasuda model [5]:
\begin{equation}
\eta(\dot{\gamma})
=
\eta_{\infty}
+
\frac{\eta_{0}-\eta_{\infty}}
{\left[1+\left(\dot{\gamma}/\dot{\gamma}_{c}\right)^{b}\right]^{(1-n)/b}}.
\label{eq:CY}
\end{equation}
For the generalized viscoelastic fluid, the fitted parameters were $\eta_0 = 4~\mathrm{Pa\cdot s}$, $\eta_\infty = 0.23~\mathrm{Pa\cdot s}$, $\dot{\gamma}_c = 0.08~\mathrm{s^{-1}}$, $n = 0.4$, and $b = 3.7$. For the Boger fluid, the fitted parameters were $\eta_0 = 0.32~\mathrm{Pa\cdot s}$, $\eta_\infty = 0.12~\mathrm{Pa\cdot s}$, $\dot{\gamma}_c = 0.3~\mathrm{s^{-1}}$, $n = 0.9$, and $b = 9$.
Extensional rheology was measured using a capillary breakup extensional rheometer (CaBER), from which the relaxation time $\lambda_0$ was extracted from the exponential thinning of the filament radius \cite{gaillard_beware_2024}. The measured relaxation times were $\lambda_0=2.7 \pm 0.05 \mathrm{~s}$ for the viscoelastic fluid and $\lambda_0=1.0 \pm 0.09 \mathrm{~s}$  for the Boger fluid.

Microfluidic devices were fabricated using standard soft lithography \cite{Xia1998}. %(Whitesides et al., Angew. Chem. Int. Ed., 2001)
Devices were molded in polydimethylsiloxane (PDMS; Sylgard 184, Dow Corning) and plasma bonded to glass microscope slides. 
The microchannels had overall dimensions of $ 240$~$\mu$m wide, 200~$\mu$m high, and 2~cm long. 
Two specific geometries were used including a single cylinder (diameter, 80~$\mu$m) centered in the channel (figure~\ref{fig:singlecylinder_SU}), and two cylinders arranged in the streamwise direction (center-to-center distance, 160~$\mu$m; figure~\ref{fig:twocylinder_SU}). 
 
Flow rates (0.02--1~$\mu$L/min) were controlled using a precision syringe pump (PhD Ultra, Harvard Apparatus) fitted with gastight glass syringes (Hamilton) to ensure stable, pulsation-free flow. Reynolds numbers across all flow conditions were $\mathrm{Re} = \rho U_{in} W /\eta(\dot{\gamma}) \le 4 \times 10^{-5}$. 
Flow was visualized by seeding the solutions with fluorescent tracer particles (diameter 0.5~$\mu$m; Thermo Fisher Scientific). Imaging was performed using an inverted microscope (Nikon Ti-E; $10\times$, 0.3 NA objective). 
Images were captured using a sensitive camera (Andor Zyla 5.5) at frame rates ranging from 12.5 to 125~frames/s. The resulting image sequences were analyzed using particle image velocimetry (PIV) implemented in MATLAB~\cite{Thielicke2014}. Velocity fields were post-processed using a local median filter for vector rejection and Gaussian spatiotemporal smoothing with (one pixel and one frame standard deviation kernel).

\section{Results and Discussion}
\label{sec:results and discussion}

Here, verification and application of the Lagrangian formulation for evaluating the polymeric stress in a viscoelastic flow are presented. First, shear flow results are compared with exact and asymptotic solutions. Next, planar channel flow results for the Lagrangian stress reconstruction are compared with Eulerian numerical results. Finally, the Lagrangian approach is applied to non-trivial, two-dimensional channel flows comprising one and two circular obstacles. 
Both numerical and experimental velocity fields are used to evaluate the stretching field and reconstruct the polymeric stress field across two different fluid rheologies and $\mathrm{Wi}$.

\subsection{Steady Shear Flow}

The FENE-P  constitutive behavior is first examined in steady shear flow, $\mathbf{u}=(u_1,u_2,u_3)=(\dot{\gamma}x_2,0,0)$, where \(\dot{\gamma}\) is the imposed constant shear rate. The corresponding deformation gradient tensor is
\begin{equation}\label{8e}
\mathbf{F}(t)=
\begin{bmatrix}
1 & 0 & 0\\
\dot{\gamma}t & 1 & 0\\
0 & 0 & 1
\end{bmatrix}.
\end{equation}
Substituting the above expression for the deformation gradient tensor into the Lagrangian formulation in equation~\ref{7e} gives the following expression for the conformation tensor \(\mathbf{A}_L(t)\):
\begin{multline}\label{9e}
\mathbf{A}_L(t)=
\begin{bmatrix}
1+\dot{\gamma}^2t^2 & \dot{\gamma}t & 0\\
\dot{\gamma}t & 1 & 0\\
0 & 0 & 1
\end{bmatrix}
\exp\!\left(
-\int_0^t
\frac{1}{\lambda_0}
f\!\left(\operatorname{tr}(\mathbf{A}_L(s))\right)\,ds
\right)
\\
+\frac{a}{\lambda_0}
\exp\!\left(
-\int_0^t
\frac{1}{\lambda_0}
f\!\left(\operatorname{tr}(\mathbf{A}_L(s))\right)\,ds
\right)
\int_{0}^{t}
\exp\!\left(
\int_0^{t'}
\frac{1}{\lambda_0}
f\!\left(\operatorname{tr}(\mathbf{A}_L(s))\right)\,ds
\right)
\\
\times
\begin{bmatrix}
1+\dot{\gamma}^2-2\dot{\gamma}^2tt'+\dot{\gamma}^2t^2 & -\dot{\gamma}t'+\dot{\gamma}t & 0\\
-\dot{\gamma}t'+\dot{\gamma}t & 1 & 0\\
0 & 0 & 1
\end{bmatrix}
\,dt'.
\end{multline}

 In the limit, \(t\rightarrow \infty\), equation~(\ref{9e}) reduces to the steady solution, \(\mathbf{A}_{L,\infty}\):
\begin{equation}\label{eq:shear_steady_AL}
\mathbf{A}_{L,\infty}
=
\begin{bmatrix}
\dfrac{a}{f_\infty}
\left[
1+2\left(\dfrac{\lambda_0\dot{\gamma}}{f_\infty}\right)^2
\right]
&
\dfrac{a\lambda_0\dot{\gamma}}{f_\infty^2}
&
0
\\[12pt]
\dfrac{a\lambda_0\dot{\gamma}}{f_\infty^2}
&
\dfrac{a}{f_\infty}
&
0
\\[12pt]
0
&
0
&
\dfrac{a}{f_\infty}
\end{bmatrix},
\end{equation}
where $f_\infty=f\!\left(\operatorname{tr}(\mathbf{A}_{L,\infty})\right).$
In the additional limit that \(L\rightarrow \infty\), where \(f_\infty=1\) and \(a=1\), equation~(\ref{eq:shear_steady_AL}) reduces to the Lagrangian Oldroyd-B steady solution:
\begin{equation}\label{eq:shear_oldroydB_AL}
\mathbf{A}_{L,\infty}^{OB}
=
\begin{bmatrix}
1+2\lambda_0^2\dot{\gamma}^2 & \lambda_0\dot{\gamma} & 0\\
\lambda_0\dot{\gamma} & 1 & 0\\
0 & 0 & 1
\end{bmatrix},
\end{equation}
which corresponds to the steady conformation tensor for the Oldroyd-B model in simple shear flow.  
% To assess the performance of the Lagrangian approach, the conformation tensor in equation~\eqref{9e} is evaluated at a large but finite time, here taken as \(t=4\lambda_0\). This value is sufficient for the transient exponential contributions to decay, so that the resulting tensor represents the steady response.
For steady shear flows, exact analytical solutions are available from Sureshkumar, Beris, and Handler~\cite{sureshkumar1997direct}, where we provide their closed-form steady-shear FENE-P tensor below for convenience:
\[
A_{11}=\frac{1}{F}\!\left(1+\frac{2\mathrm{Wi}^{2}}{F^{2}}\right), \qquad
A_{22}=A_{33}=\frac{1}{F}, \qquad
A_{12}=\frac{\mathrm{Wi}}{F^{2}}.
\]
Where $\mathrm{Wi}=\lambda_0\dot{\gamma}$. The auxiliary quantities are given by
\[
F=\frac{\sqrt{3}\,\Omega}{2\sinh(\phi/3)}, \qquad
\phi=\sinh^{-1}\!\left(\frac{3\sqrt{3}}{2}\Omega\right), \qquad
\Omega=\frac{\sqrt{2}\,\mathrm{Wi}}{L}.
\]
Additionally, asymptotic theory is also available in the small- and large-Weissenberg-number limit from the expansion of Yamani and McKinley~\cite{yamani2023master}. For \(\mathrm{Wi}\ll 1\) and \(\mathrm{Wi}\gg 1\), the asymptotic solutions are 
\[
\operatorname{tr}(\mathbf{A})=3+2\mathrm{Wi}^{2},
\]
and
\[
\operatorname{tr}(\mathbf{A}) = 
L^{2}-(L^{2}-3)\left(\frac{L}{\sqrt{2}\,\mathrm{Wi}}\right)^{2/3},
\]
respectively.

\begin{figure*}[!t]
    \centering
    % Top large figure
    \includegraphics[width=0.5\textwidth, trim={0cm 0cm 0cm 0cm},clip]{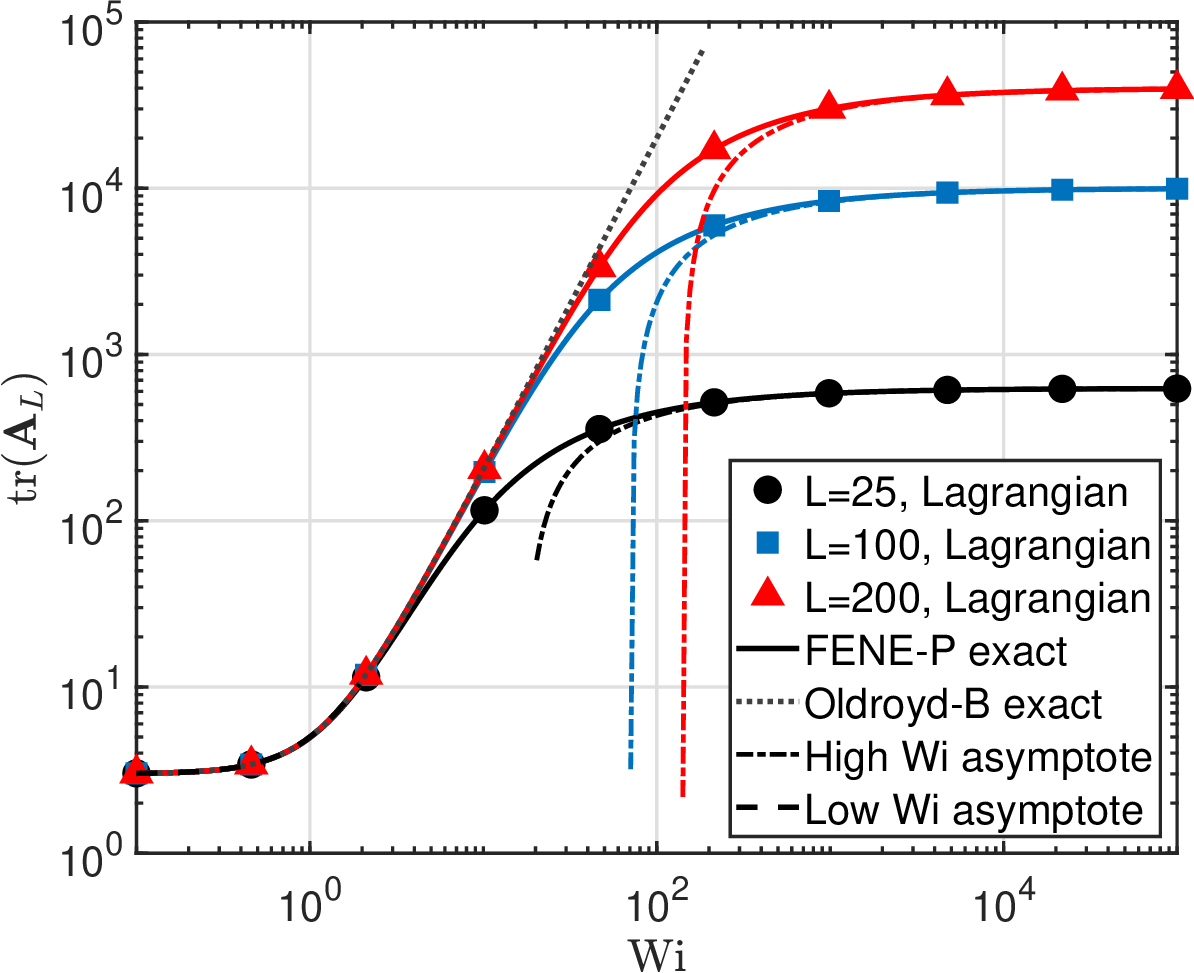}
        % Adjust vertical spacing between images
\caption{Lagrangian results for \(\operatorname{tr}(\mathbf{A}_L)\) in a steady shear flow compared with the steady FENE-P exact and asymptotic solutions, as well as the Oldroyd-B exact solution. Exact and high-\(Wi\) FENE-P solutions are also compared for three different extensibility parameters \(L\).}
    \label{fig:trvswi_shear}
\end{figure*}

 The Lagrangian FENE-P results show excellent agreement with the steady FENE-P exact solution and the asymptotic limits for different extensibility parameters (figure~\ref{fig:trvswi_shear}). 
The trace of the conformation tensor, \(\operatorname{tr}(\mathbf{A}_L)\), provides a simple metric to compare across the various solutions and is in line with typical conventions.
The Lagrangian results closely follow the steady FENE-P solution over the entire range of \(\mathrm{Wi}\). 
The low-\(\mathrm{Wi}\) behavior also agrees with corresponding asymptotic solution and  the Oldroyd-B solution, as expected. 
At larger \(\mathrm{Wi}\), the FENE-P response approaches a finite plateau set by the extensibility parameter \(L\). 
The haulted growth of the FENE-P trace at large \(\mathrm{Wi}\) shows the effect of finite extensibility, which physically corresponds to the polymer elongation approaching its upper limit. 
Conversely, the Oldroyd-B trace departs from FENE-P solutions at moderate \(\mathrm{Wi}\), owing to its unbounded Hookean spring behavior.

 \subsection{Planar Channel Flow} 
 \label{sec:planar channel flow}

 The Lagrangian formulation of the polymeric stress was next evaluated for a planar channel flow having inhomogeneous shear. In this case, the velocity field is unidirectional, but the shear rate varies across the channel width. This case provides a simple, yet non-trivial test for comparing the reconstructed stress distribution from a numerically determined velocity field, including the Eulerian reference solution for the polymeric stress tensor. The channel width is \(W=240~\mu\mathrm{m}\), and the channel length is \(L_c=10W\). The comparison is performed for \(\mathrm{Wi}=0.1\) and \(1.0\), where the channel width is used as the characteristic length scale, \(\mathrm{Wi}=\lambda_0 U_{in}/W\). 
 Both FENE-P and Oldroyd-B model fluids are considered using the same flow and material parameters, which allows the difference in the stress profiles to be associated with the constitutive model. 
The material properties are kept the same for both constitutive models, with \(\lambda_0=1\), \(\eta_p=0.95\), and \(\eta_s=0.05\). For the FENE-P model, the extensibility parameter is set to \(L^2=1000\).
 For this unidirectional channel flow, the deformation gradient is obtained from
\[
\frac{d\mathbf{F}}{dt}=\mathbf{F}\nabla\mathbf{u},
\qquad 
\mathbf{F}(0)=\mathbf{I}.
\]
Since \(\mathbf{u}=(u_1(x_2),0,0)\), the deformation gradient tensor becomes
\[
\mathbf{F}(t)=
\begin{bmatrix}
1 & 0 & 0\\
t\,\dfrac{\partial u_1}{\partial x_2} & 1 & 0\\
0 & 0 & 1
\end{bmatrix},
\]
where \(\partial u_1/\partial x_2\) is evaluated from the numerically-obtained velocity field.

\begin{figure*}[!t]
\centering

\begin{subfigure}[t]{0.35\textwidth}
    \centering
    \includegraphics[width=\textwidth, trim={0cm 0cm 0cm 0cm}, clip]
    {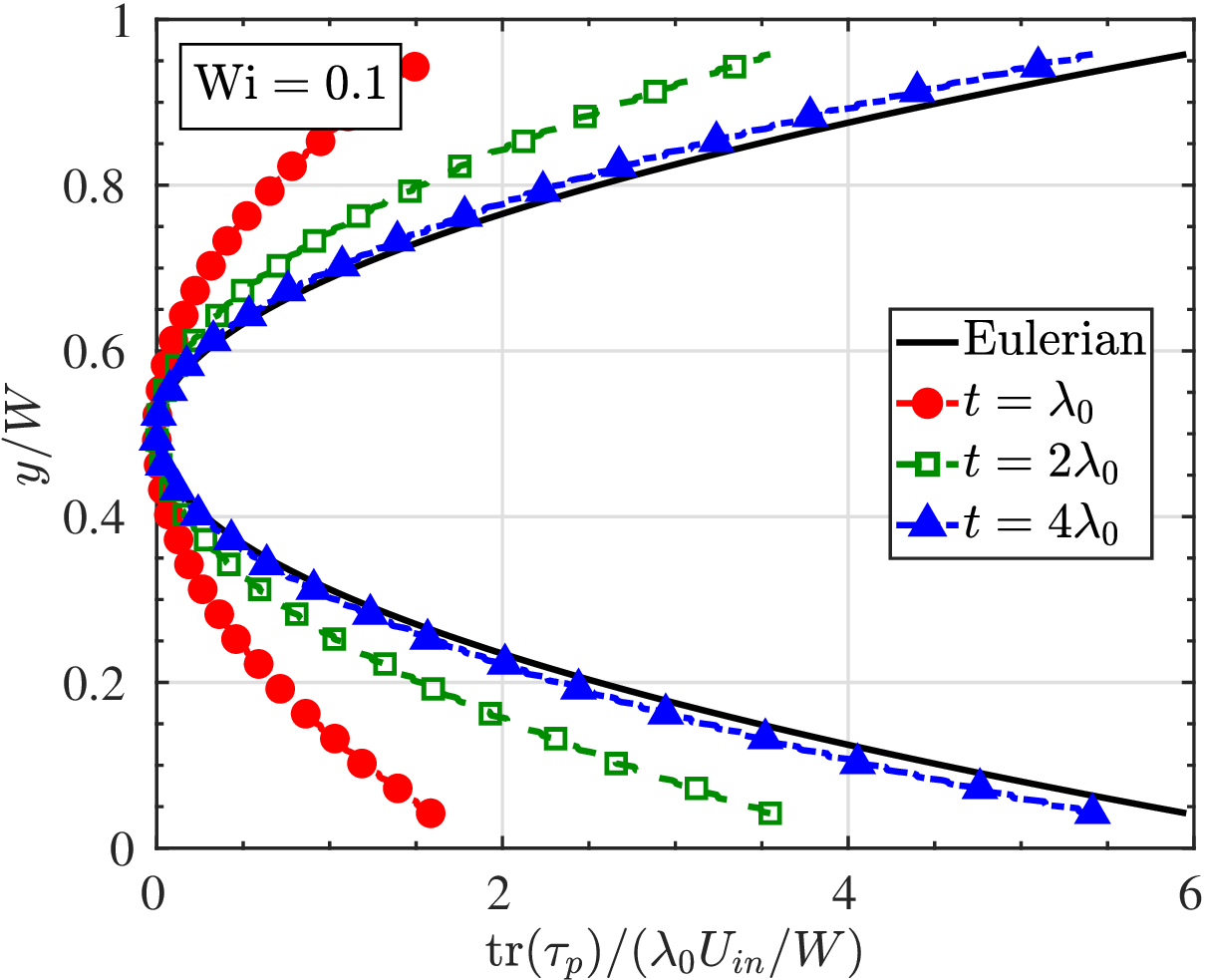}
    \caption{}
    \label{fig:planar_channel_wi0p1}
\end{subfigure}
\hspace{0.05\textwidth}
\begin{subfigure}[t]{0.35\textwidth}
    \centering
    \includegraphics[width=\textwidth, trim={0cm 0cm 0cm 0cm}, clip]
    {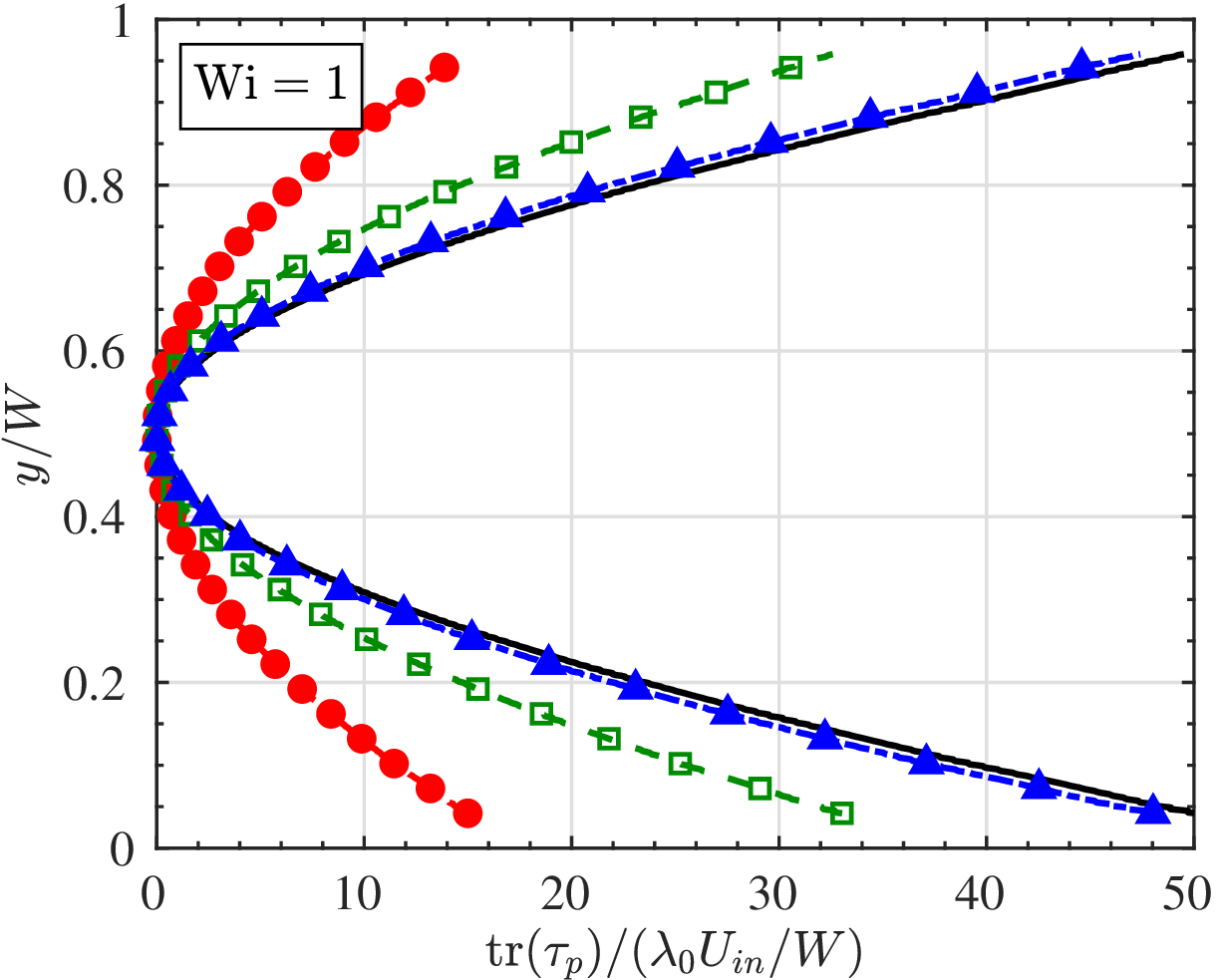}
    \caption{}
    \label{fig:planar_channel_wi1}
\end{subfigure}

\caption{Convergence of the Lagrangian reconstruction of the dimensionless trace of the polymeric stress tensor, $\mathrm{tr}(\boldsymbol{\tau}_p)/(\lambda_0 U_{in}/R)$, across a planar channel flow for the FENE-P model. Results are shown for (a) $\mathrm{Wi}=0.1$ and (b) $\mathrm{Wi}=1$.}
\label{fig:convergence}
\end{figure*}

A comparison of the Lagrangian and Eulerian results is made using the profile of the trace of the polymeric stress, \(\operatorname{tr}(\boldsymbol{\tau}_p)\), across the channel width at a fixed streamwise location where the velocity field has already  reached a fully developed profile.   For both $\mathrm{Wi}=0.1$ and $\mathrm{Wi}=1$, the FENE--P Lagrangian stress reconstruction converges to the Eulerian reference solution as the integration time increases from $t=\lambda_0$ to $t=4\lambda_0$ (figure~\ref{fig:convergence}). 

\begin{figure*}[!t]
\centering

\begin{subfigure}[t]{0.35\textwidth}
    \centering
    \includegraphics[width=\textwidth, trim={0cm 0cm 0cm 0cm}, clip]
    {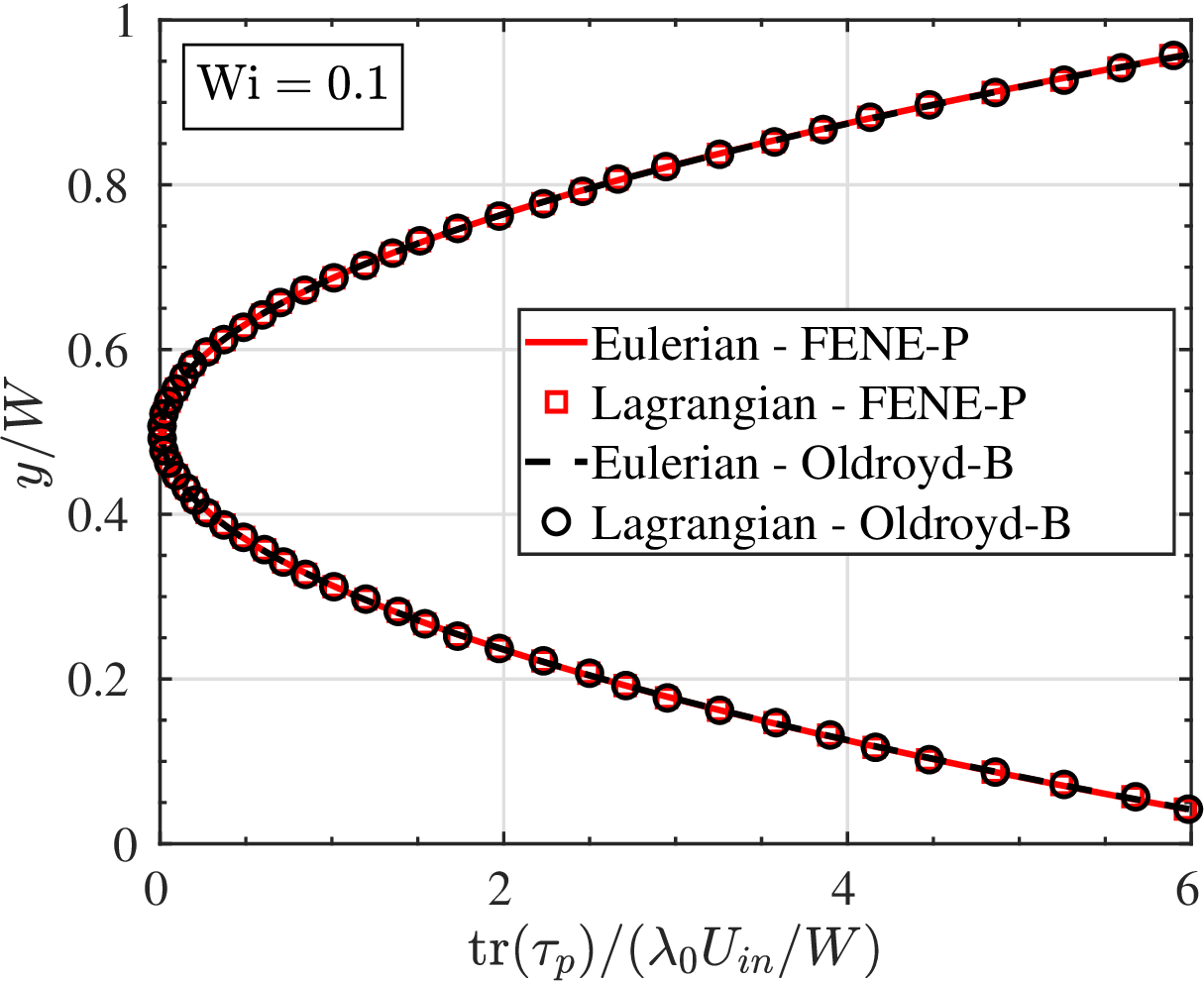}
    \caption{}
    \label{fig:planar_channel_fenep_oldb_wi0p1}
\end{subfigure}
\hspace{0.05\textwidth}
\begin{subfigure}[t]{0.35\textwidth}
    \centering
    \includegraphics[width=\textwidth, trim={0cm 0cm 0cm 0cm}, clip]
    {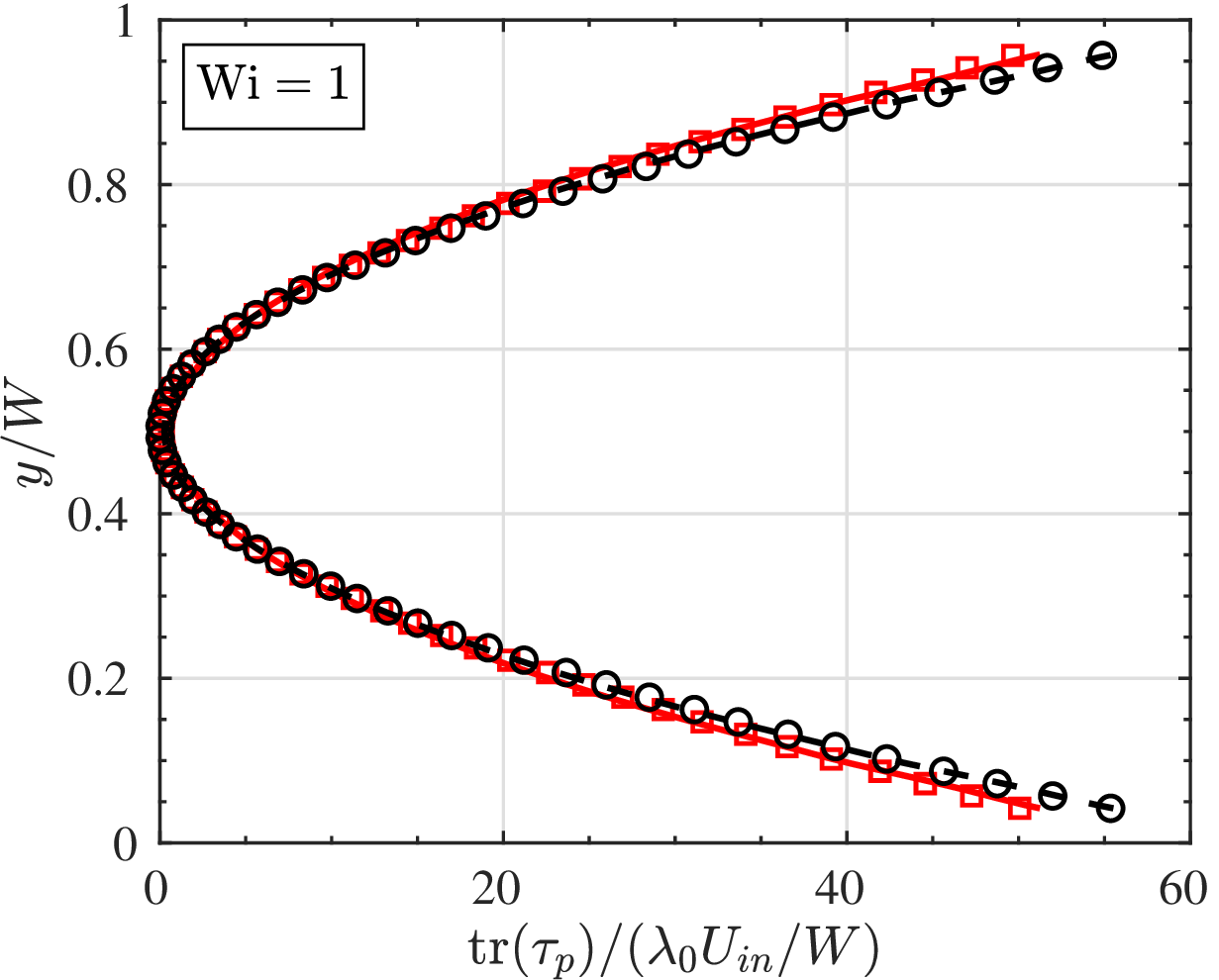}
    \caption{}
    \label{fig:planar_channel_fenep_oldb_wi1}
\end{subfigure}

\caption{Comparison of the dimensionless polymeric-stress trace, $\mathrm{tr}(\boldsymbol{\tau}_p)/(\lambda_0 U_{in}/W)$, across planar channel flow for the FENE-P and Oldroyd-B models. The Lagrangian stress reconstruction obtained from the velocity field is compared with the Eulerian reference solution from RheoTool. Results are shown for (a) $\mathrm{Wi}=0.1$ and (b) $\mathrm{Wi}=1$.}
\label{fig:planar_channel_fenep_oldb}
\end{figure*}

Figure~\ref{fig:planar_channel_fenep_oldb} shows the results for the FENE-P and Oldroyd-B models at these two different \(\mathrm{Wi}\) numbers. The Lagrangian stress profiles shown in this comparison are reconstructed using an integration time of $t = 4\lambda_0$. In each case, the Lagrangian stress reconstructed from the numerical velocity field is compared with the Eulerian numerical stress field, which is used as the reference solution. For both constitutive models, there is excellent agreement between the reconstructed stress from the Lagrangian formulation and the Eulerian reference solution. The \(L^2\)-norm error between the Lagrangian reconstruction and the Eulerian numerical result is between \(1\%\) and \(2\%\).
As \(\mathrm{Wi}\) increases, the Oldroyd-B and FENE-P profiles begin to separate, consistent with the steady shear flow results in figure ~\ref{fig:trvswi_shear} where the FENE-P trace was limited by finite extensibility while the Oldroyd-B trace continued to grow. This behavior further motivates the idea of extending the Lagrangian stress reconstruction beyond the Oldroyd-B limit to the FENE-P model.

 \subsection{Channel Flows with Obstacles} 
 
%We next evaluate the Lagrangian formulation in channel flows with obstacles. 
Expanding upon the unidirectional planar channel flow (section~\ref{sec:planar channel flow}), arrangements of circular cylinders were incorporated in the center of channel to demonstrate the Lagrangian reconstruction of polymeric stress in nontrivial geometries. In addition to strong shear, these mixed kinematic, two-dimensional flows introduce strongly extensional regions near the stagnation points on the cylinders, which generate high polymeric stress. The polymeric stress reconstruction is performed for velocity field data obtained from both microfluidic experiments and matching numerical simulations. Two geometries are considered: (\textit{i}) a single circular cylinder placed in the center of the channel and (\textit{ii}) two identical circular cylinders aligned in the streamwise direction. Each cylinder has radius \(r = 40~\mu\mathrm{m}\). For the two-cylinder case, the center-to-center distance between the cylinders is \(d = 4r\). 

To match with microfluidic experiments, the FENE-P case corresponds to \(\mathrm{Wi}=0.68\) with \(\lambda_0=2.7\mathrm{~s}\), and the Oldroyd-B case corresponds to \(\mathrm{Wi}=0.25\) with \(\lambda_0=1.0\mathrm{~s}\).
These different \(\mathrm{Wi}\) values arise from experiments with the same inlet flow speed but having two different fluids with different relaxation times. In addition, for the single-cylinder case, the Oldroyd-B model was also analyzed at the same \(\mathrm{Wi}=0.68\) to provide a direct comparison with the FENE-P model at the same elastic strength.   The material properties are kept the same as those introduced above (section~\ref{sec:planar channel flow}).
To match the experimental Boger fluid used for the Oldroyd-B case (section~\ref{sec:experimental methods}), the solvent viscosity and the polymeric viscosity are set to  \(\eta_s=0.36 \mathrm{~Pa. s}\) and \(\eta_p=0.64 \mathrm{~Pa. s}\), respectively.

 \subsubsection{Velocity and Stretching Fields} 
 
\begin{figure*}[!t]
    \centering
    % Top large figure
    \includegraphics[width=0.75\textwidth, trim={0cm 0cm 0cm 0cm},clip]{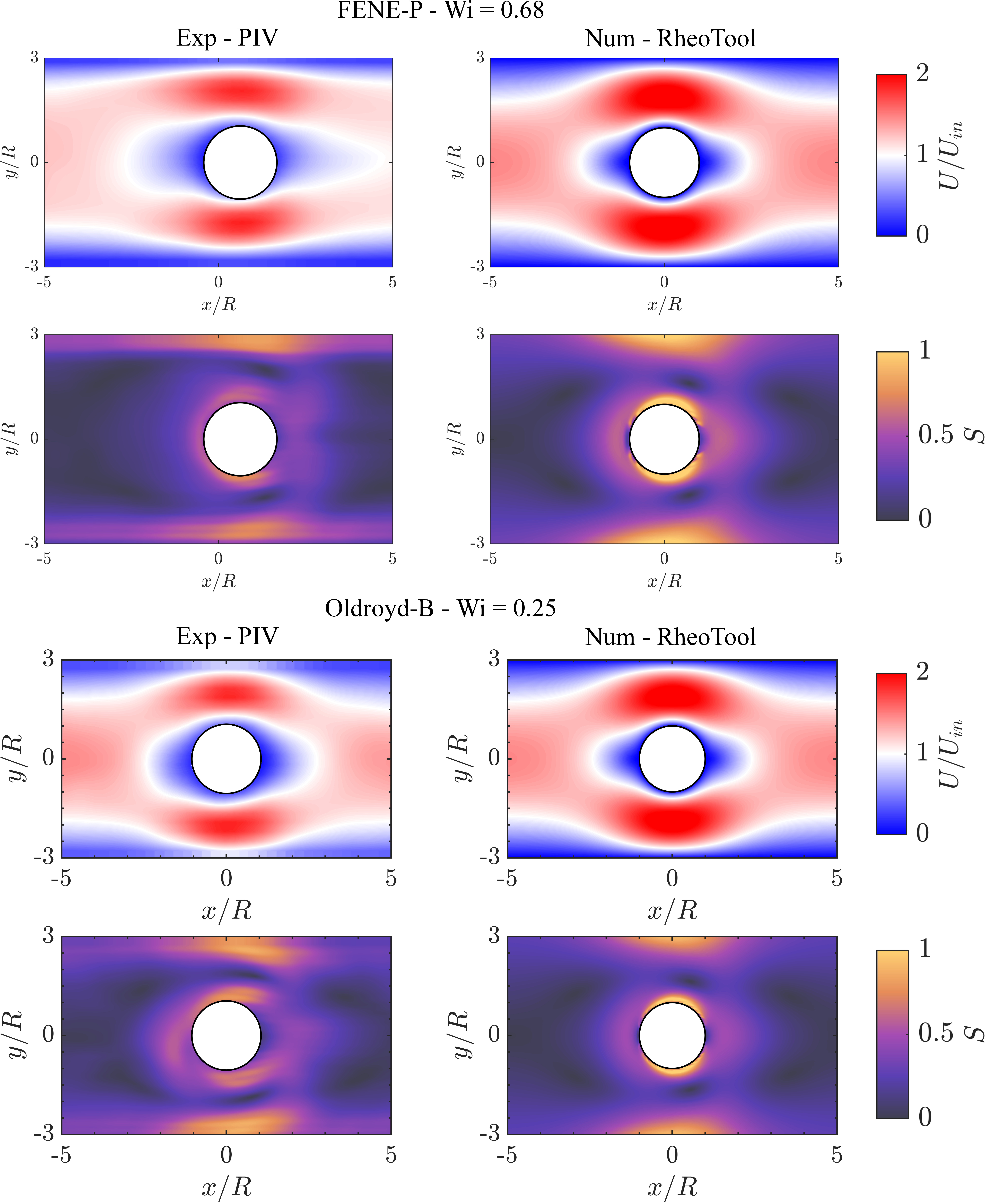}
        % Adjust vertical spacing between images
 \caption{Comparison of the non-dimensional velocity magnitude and stretching field for flow past one circular cylinder. The FENE-P case corresponds to \(\mathrm{Wi}=0.68\), and the Oldroyd-B case corresponds to \(\mathrm{Wi}=0.25\). For each model, the first row shows the non-dimensional velocity field \(U/U_{in}\), and the second row shows the stretching field \(S\). Experimental results are shown in the left column and simulation results are shown in the right column.}
    \label{fig:singlecylinder_SU}
\end{figure*}

\begin{figure*}[!t]
    \centering
    % Top large figure
    \includegraphics[width=0.75\textwidth, trim={0cm 0cm 0cm 0cm},clip]{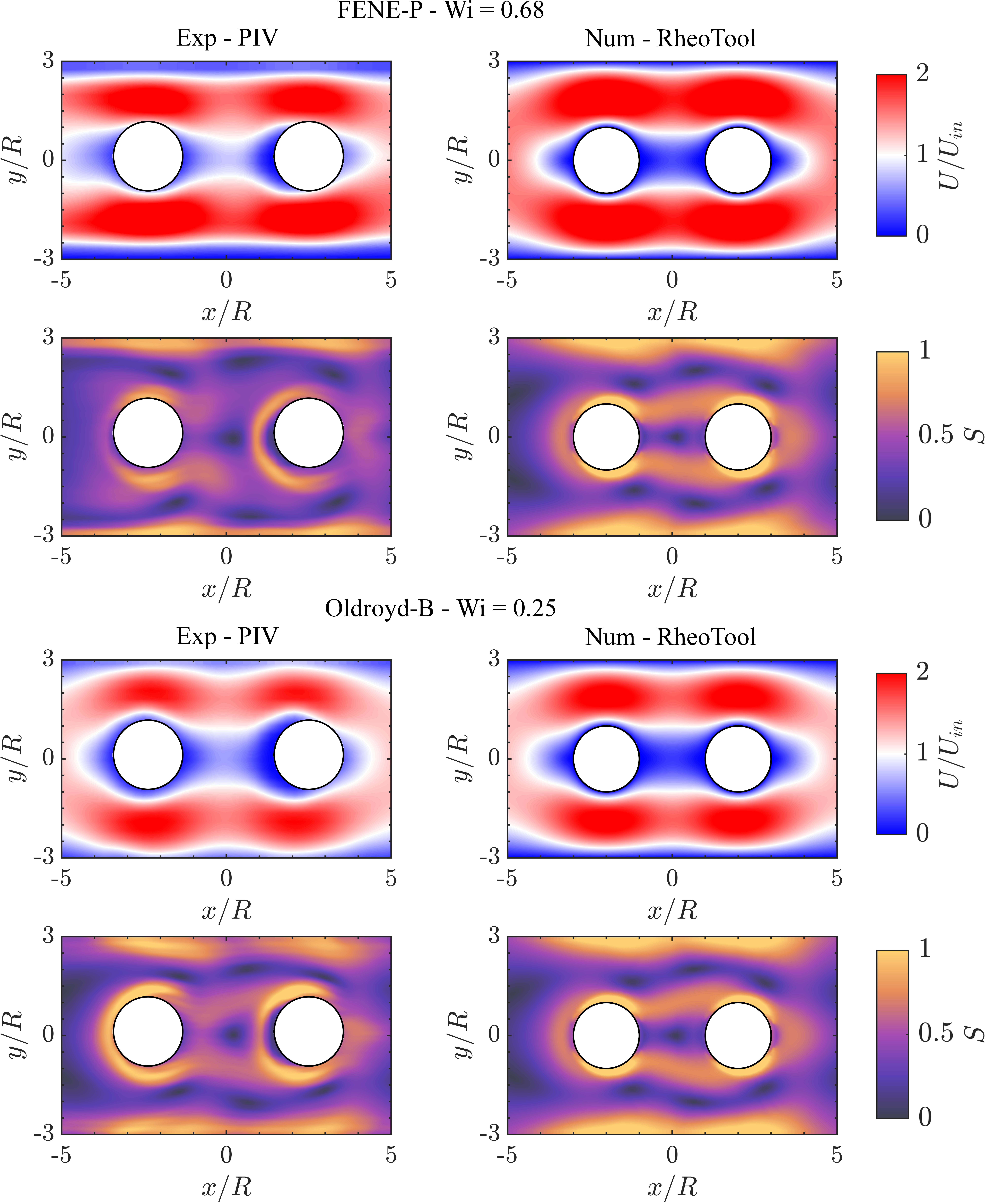}
        % Adjust vertical spacing between images
\caption{Comparison of the non-dimensional velocity magnitude and stretching field for flow through a channel with two circular cylinders. The FENE-P case corresponds to \(\mathrm{Wi}=0.68\), and the Oldroyd-B case corresponds to \(\mathrm{Wi}=0.25\). For each model, the first row shows the non-dimensional velocity field \(U/U_{in}\), and the second row shows the stretching field \(S\). Experimental results are shown in the left column and simulation results are shown in the right column.}
    \label{fig:twocylinder_SU}
\end{figure*}

Velocity and Lagrangian stretching fields are first examined to compare the simulated and experimental flow fields and their resulting deformation patterns. The experimental velocity field is obtained from PIV measurements in the microfluidic devices, while simulated velocity fields are obtained from numerical integration of Eulerian equations.  
Figures~\ref{fig:singlecylinder_SU} and~\ref{fig:twocylinder_SU} compare the non-dimensional velocity magnitude and stretching field for the one-cylinder and two-cylinder geometries, respectively. In both cases, the simulated and experimental velocity fields show similar flow structures around the cylinders. Although the velocity fields are steady, a fluid particle moving around the cylinders still experiences a Lagrangian unsteady deformation history along its trajectory. This accumulated deformation is reflected in the stretching fields.

In the present cases, the stretching fields from the simulated and experimental velocity fields show similar high-stretching regions near the cylinders and, for the two-cylinder case, in the gap between the cylinders. The Lagrangian fluid stretching is evaluated for an integration time of 2\( \lambda_0\): $2.0 \mathrm{~s}$ for the Oldroyd-B model and $5.4 \mathrm{~s}$ for the FENE-P models, which is consistent with polymeric stress reconstruction below (see section~\ref{sec:polymeric stress reconstruction}).  Regions of high stretching are also where stronger polymeric stress is expected, consistent with the connection between Lagrangian stretching and stress topology discussed earlier. 
While the stretching field is highly correlated with stress topology, it cannot directly provide the value of the constitutive stress without incorporating a rheological model.  For this reason, we demonstrate the utility of the present Lagrangian approach in determining the polymeric stress tensor for these geometries in the section below.

 \subsubsection{Polymeric Stress Reconstruction} 
 \label{sec:polymeric stress reconstruction}

\begin{figure*}[!t]
    \centering
    % Top large figure
    \includegraphics[width=0.9\textwidth, trim={0cm 0cm 0cm 0cm},clip]{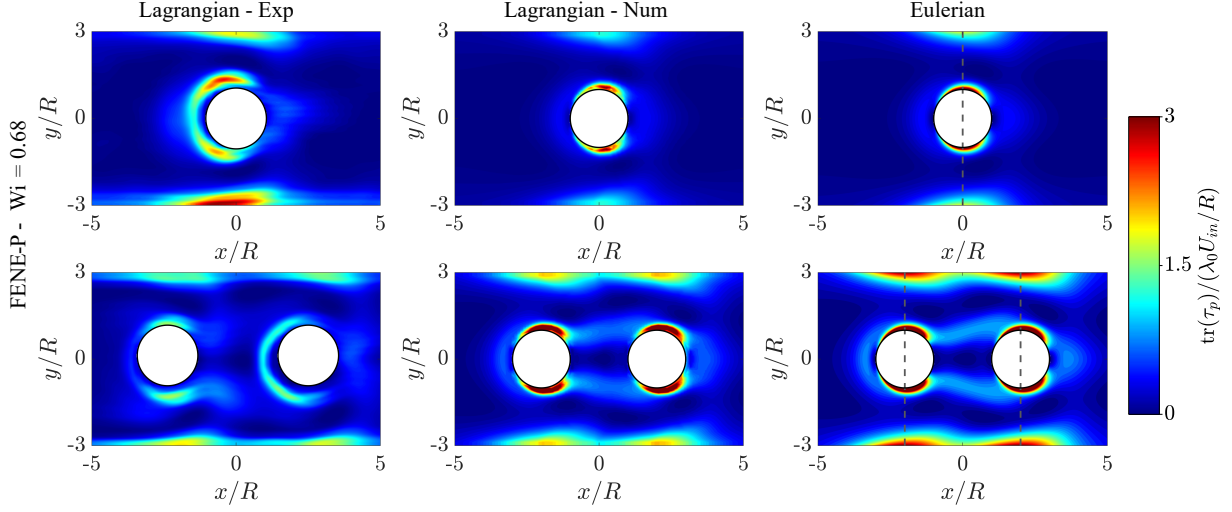}
        % Adjust vertical spacing between images
\caption{Comparison of the non-dimensionalized trace of the polymeric stress for the FENE-P model at   \(\mathrm{Wi}=0.68\). Results are shown for flow past one circular cylinder in the top row and two circular cylinders in the bottom row. The first column shows the Lagrangian reconstruction using the experimental velocity field (Lagrangian - Exp), the second column illustrates the Lagrangian reconstruction using the numerical velocity field (Lagrangian - Num), and the third column demonstrates the Eulerian numerical reference solution. The dashed vertical lines indicate locations for profiles of the polymeric-stress trace (see figure~\ref{fig:fenep_profile_comparison}).}
    \label{fig:fenep_trTau_onevstwo}
\end{figure*}

\begin{figure*}[!t]
    \centering
    % Top large figure
    \includegraphics[width=0.9\textwidth, trim={0cm 0cm 0cm 0cm},clip]{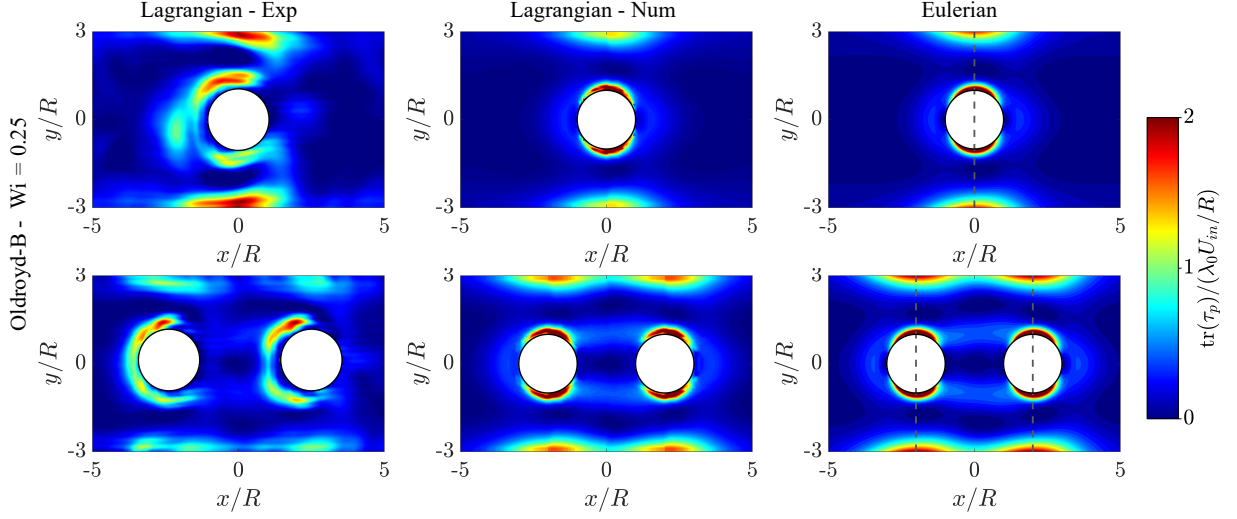}
        % Adjust vertical spacing between images
\caption{Comparison of the non-dimensional polymeric-stress trace for the Oldroyd-B model at  \(\mathrm{Wi}=0.25\). Results are shown for flow past one circular cylinder in the top row  and two circular cylinders in the bottom row. The first column shows the Lagrangian reconstruction using the experimental velocity field (Lagrangian - Exp), the second column illustrates the Lagrangian reconstruction using the numerical velocity field (Lagrangian - Num), and the third column demonstrates the Eulerian numerical reference solution. 
The dashed vertical lines indicate the centreline locations for profiles of the polymeric-stress trace (see figure~\ref{fig:oldB_profile_comparison}).}
    \label{fig:oldB_trTau_onevstwo}
\end{figure*}

 The dimensionless trace of the polymeric stress tensor is shown in figures~\ref{fig:fenep_trTau_onevstwo} and~\ref{fig:oldB_trTau_onevstwo} for the FENE-P and Oldroyd-B models, respectively. The Lagrangian results obtained from the experimental and numerical velocity fields are compared with the Eulerian numerical reference solution.   For both the FENE-P and Oldroyd-B models, the Lagrangian reconstruction based on the numerical velocity field shows good agreement with the Eulerian reference solution, with only minor differences. The Lagrangian reconstruction based on the experimental velocity field also captures the main stress patterns and their magnitude around the circular obstacles. Compared with the stretching field, which only shows the regions of high accumulated deformation, the Lagrangian approach also provides the polymeric-stress magnitude and includes the effects of polymer relaxation and finite extensibility. Therefore, the stretching field \cite{kumar2023lagrangian} alone does not uniquely determine the polymeric-stress magnitude, particularly when comparing different constitutive models.

For the Lagrangian stress reconstruction, the integration time was $t=2\lambda_0$, which is consistent with the stretching analysis. An advantage of the Lagrangian approach for these steady flows is that the initial polymeric conformation is not needed for long integration times. However, long integration times require a large spatial extent of the velocity field to ensure that Lagrangian particles remain within the domain. Therefore, this shorter integration time has a tradeoff with accuracy, but still captures the main features and magnitude of the polymeric stress as illustrated below.

\begin{figure*}[!t]
    \centering

    % ===================== Row 1: trTau vs y =====================
    \begin{subfigure}[t]{0.32\textwidth}
        \centering
        \includegraphics[width=\textwidth, trim={0cm 0cm 0cm 0cm}, clip]
        {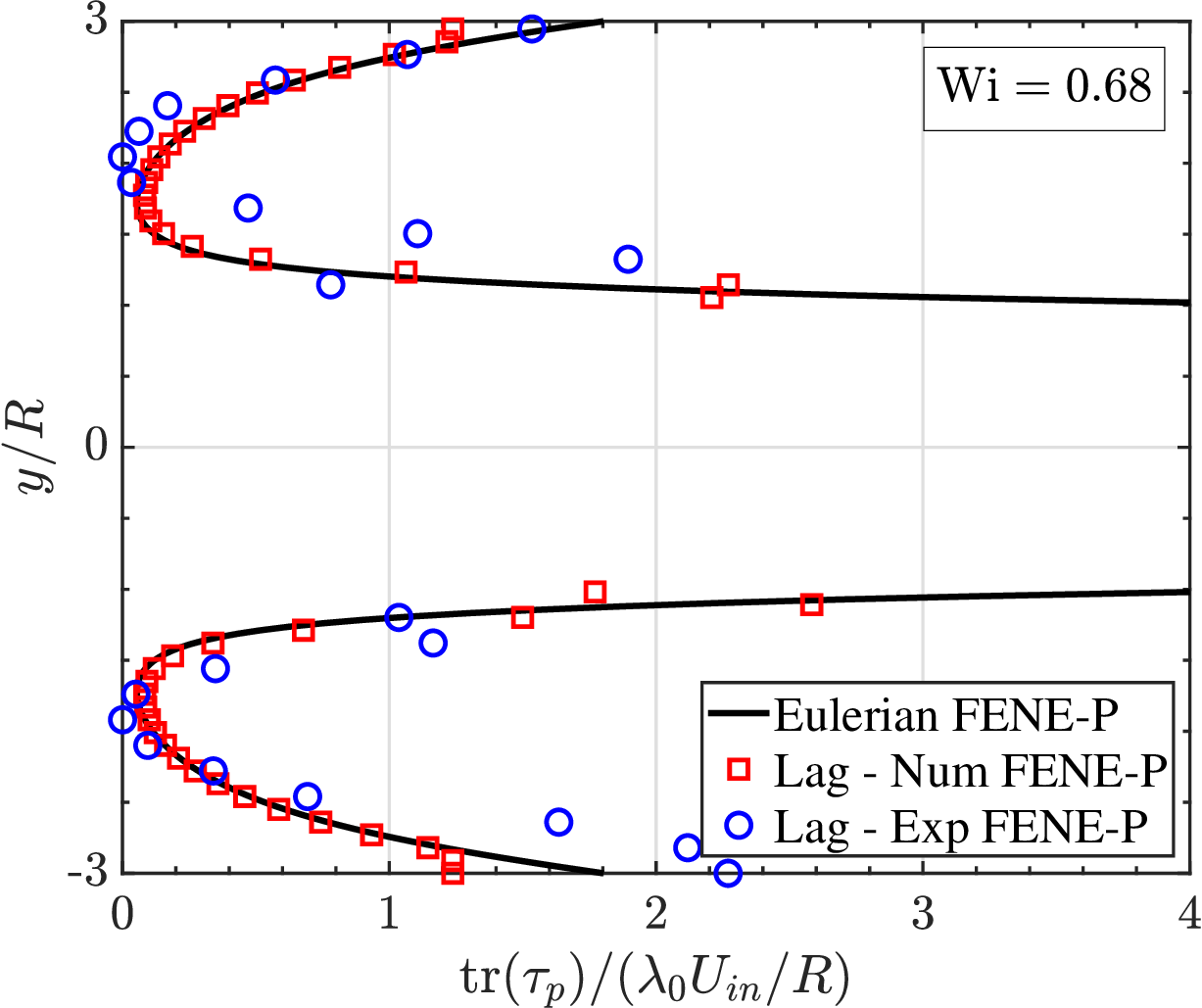}
        \caption{One cylinder}
        \label{fig:fenep_line_one_cylinder}
    \end{subfigure}
    \hfill
    \begin{subfigure}[t]{0.32\textwidth}
        \centering
        \includegraphics[width=\textwidth, trim={0cm 0cm 0cm 0cm}, clip]
        {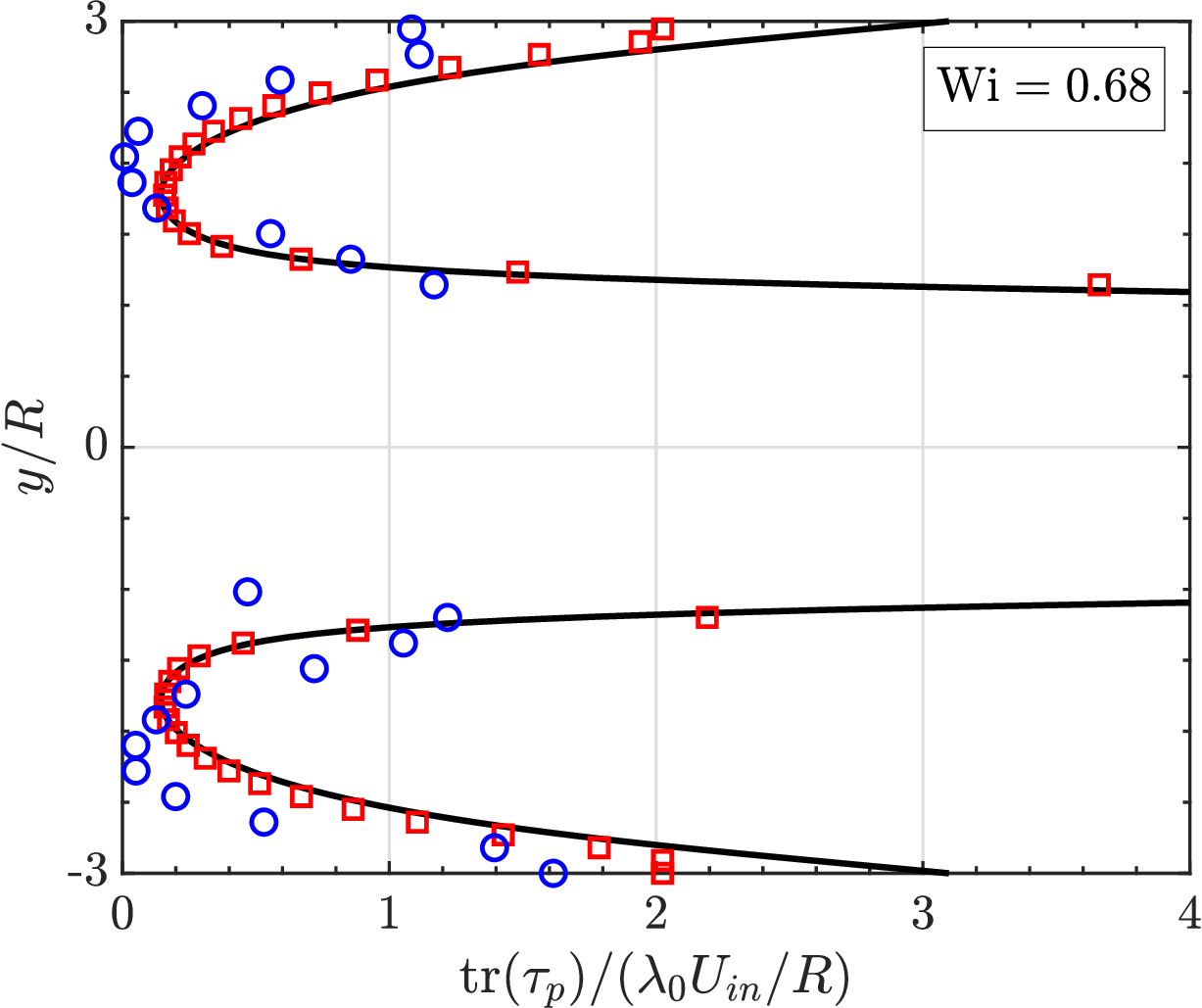}
        \caption{Two cylinders, left}
        \label{fig:fenep_line_LC}
    \end{subfigure}
    \hfill
    \begin{subfigure}[t]{0.32\textwidth}
        \centering
        \includegraphics[width=\textwidth, trim={0cm 0cm 0cm 0cm}, clip]
        {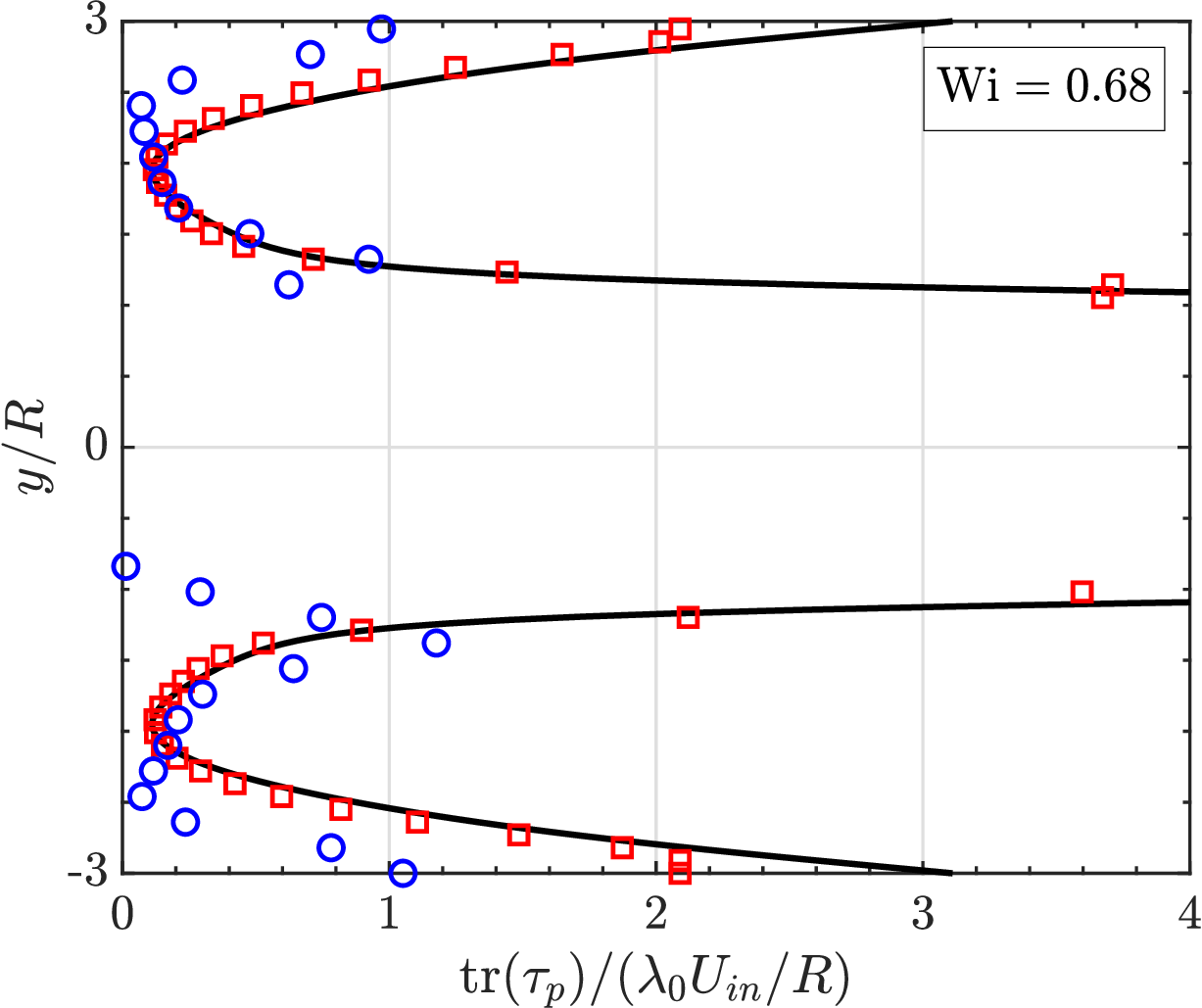}
        \caption{Two cylinders, right}
        \label{fig:fenep_line_RC}
    \end{subfigure}

    \vspace{0.15cm}

    % ===================== Row 2: circular profiles =====================
    \begin{subfigure}[t]{0.32\textwidth}
        \centering
        \includegraphics[width=\textwidth, trim={0cm 0cm 0cm 0cm}, clip]
        {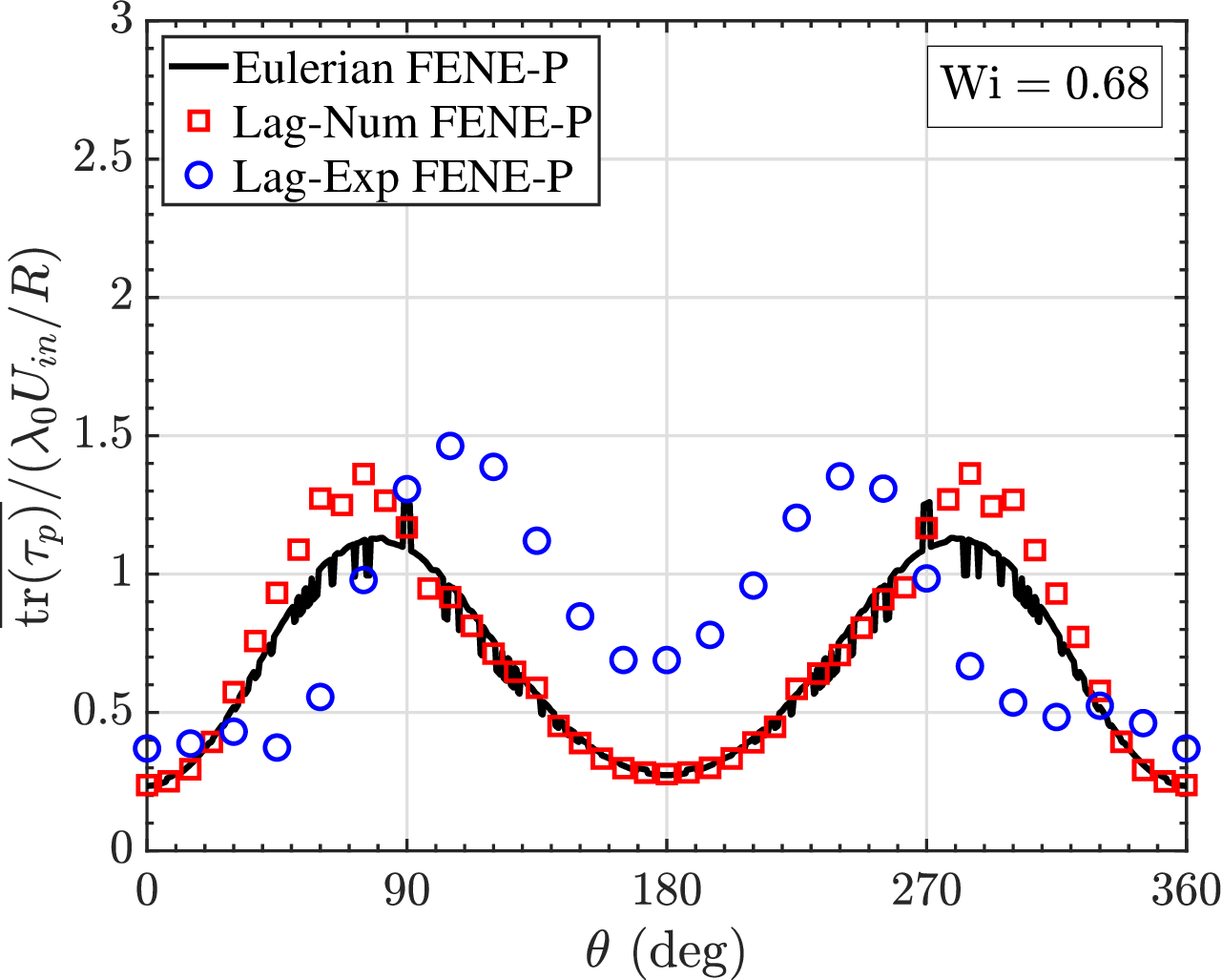}
        \caption{One cylinder}
        \label{fig:fenep_circle_one_cylinder}
    \end{subfigure}
    \hfill
    \begin{subfigure}[t]{0.32\textwidth}
        \centering
        \includegraphics[width=\textwidth, trim={0cm 0cm 0cm 0cm}, clip]
         {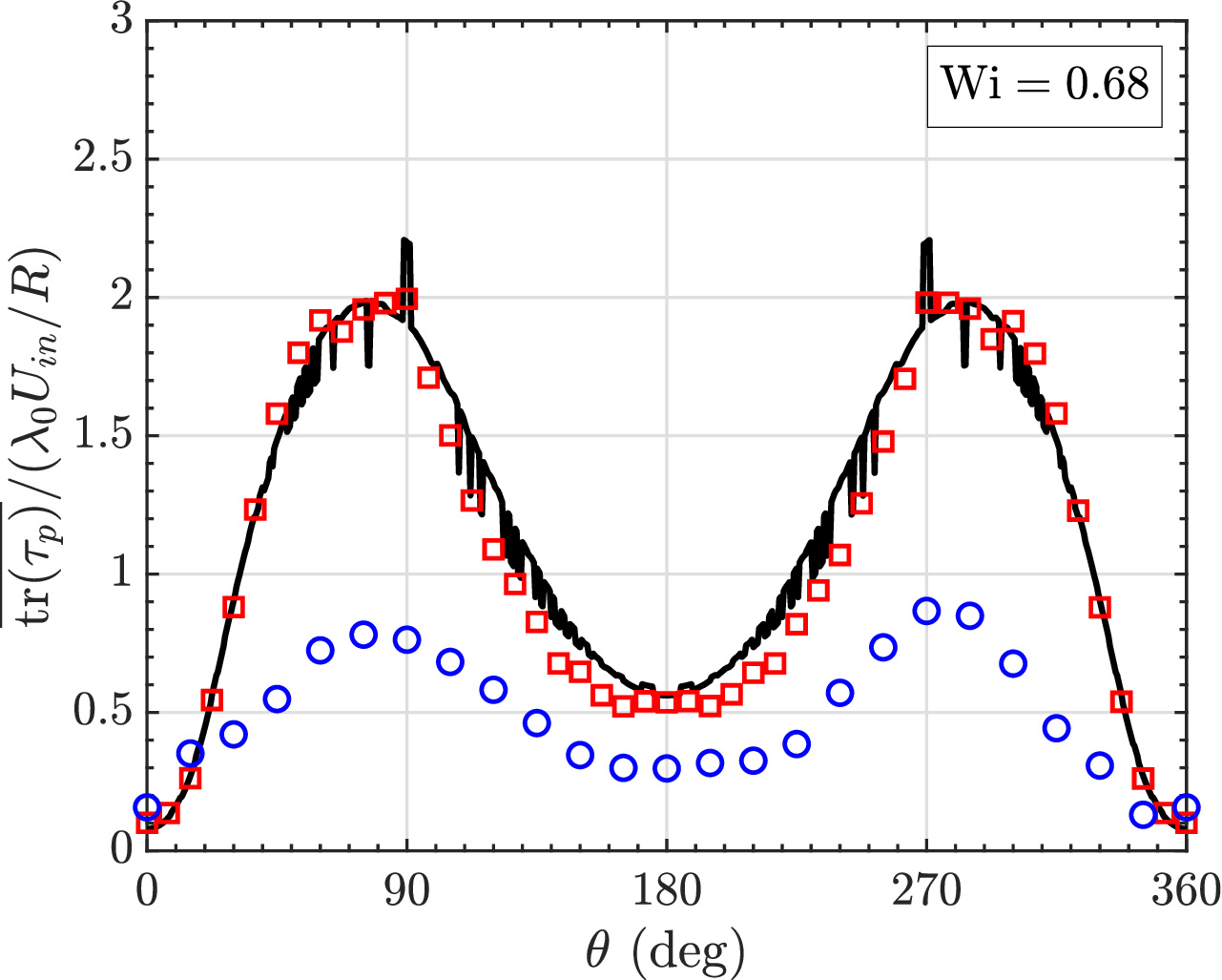}
        \caption{Two cylinders, left}
        \label{fig:fenep_circle_LC}
    \end{subfigure}
    \hfill
    \begin{subfigure}[t]{0.32\textwidth}
        \centering
        \includegraphics[width=\textwidth, trim={0cm 0cm 0cm 0cm}, clip]
        {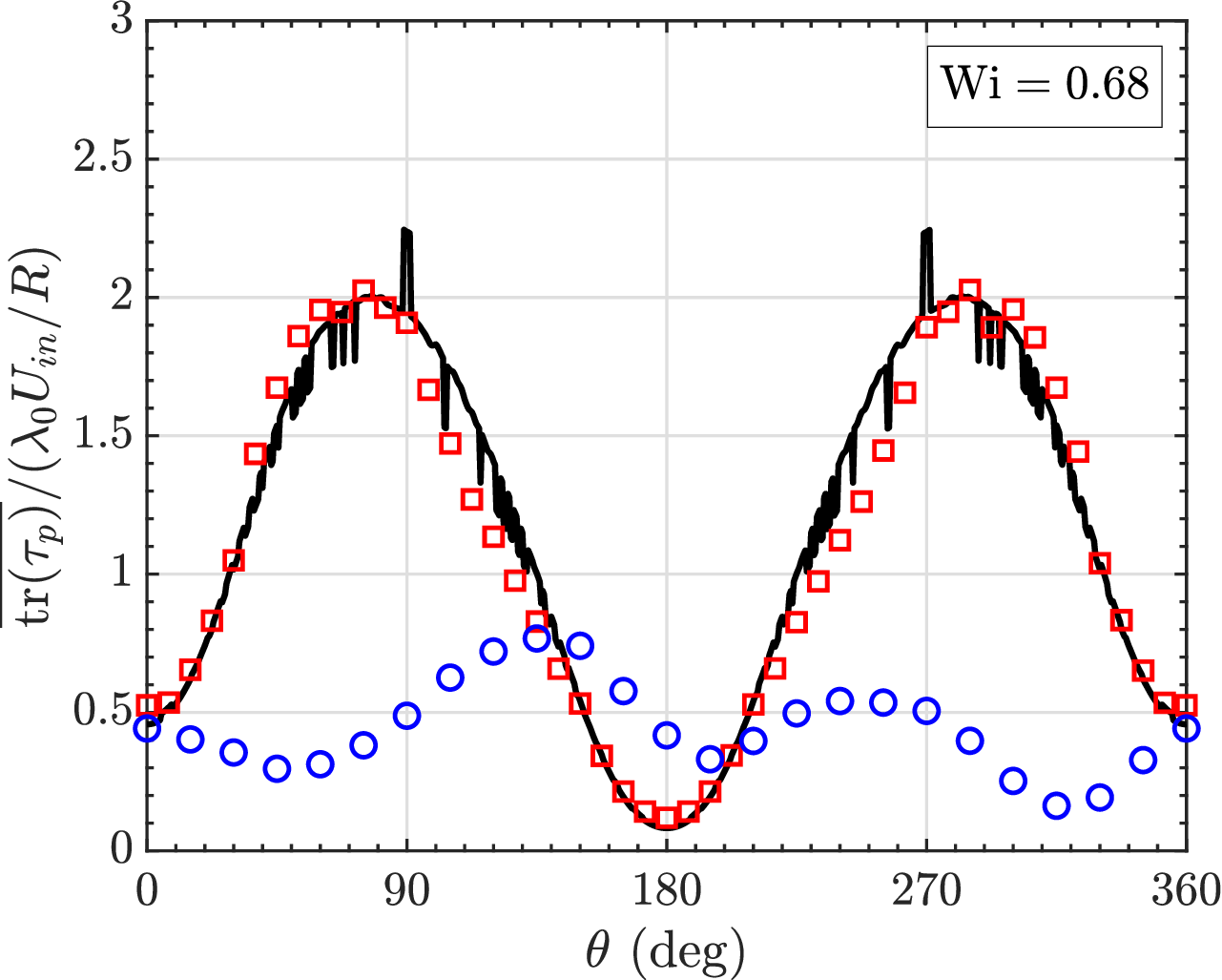}
        \caption{Two cylinders, right}
        \label{fig:fenep_circle_RC}
    \end{subfigure}
\caption{Profile comparison (see figure~\ref{fig:fenep_trTau_onevstwo}) of the non-dimensional polymeric-stress trace for the FENE-P model at $\mathrm{Wi}=0.68$. Red squares show the Lagrangian reconstruction using the numerical velocity field (Lag-Num), blue circles show the Lagrangian reconstruction using the experimental velocity field (Lag-Exp), and black lines show the Eulerian numerical reference solution. The top row shows $\mathrm{tr}(\boldsymbol{\tau}_p)$ profiles across $y/R$, while the bottom row shows radially-averaged azimuthal profiles around each cylinder over $r=40$--$60~\mu\mathrm{m}$. The first column corresponds to the one-cylinder case, and the second and third columns correspond to the left and right cylinders in the two-cylinder case. }
    \label{fig:fenep_profile_comparison}
\end{figure*}

 Some  differences are more visible in the Lagrangian reconstruction based on the experimental velocity field. The PIV velocity field has lower spatial resolution and greater measurement uncertainty than the simulated velocity field. These limitations introduce some inaccuracy in the calculation of the deformation-gradient history over time when tracking fluid particles in the channel. Furthermore, the locations of the microfabricated circular cylinders are slightly different from those used in the numerical geometry. However, the FENE-P and Oldroyd-B models are simplified representations of real polymer solutions and do not capture effects such as distributions of polymer sizes and relaxation times \cite{entov1997effect}, possible entanglement, and complex molecular conformations \cite{boyko2024perspective}. Despite these noted differences, the stress field topology and magnitude around the circular obstacles remain highly consistent with both the numerical Lagrangian and Eulerian results.

\begin{figure*}[!t]
    \centering

    % ===================== Row 1: trTau vs y =====================
    \begin{subfigure}[t]{0.32\textwidth}
        \centering
        \includegraphics[width=\textwidth, trim={0cm 0cm 0cm 0cm}, clip]
        {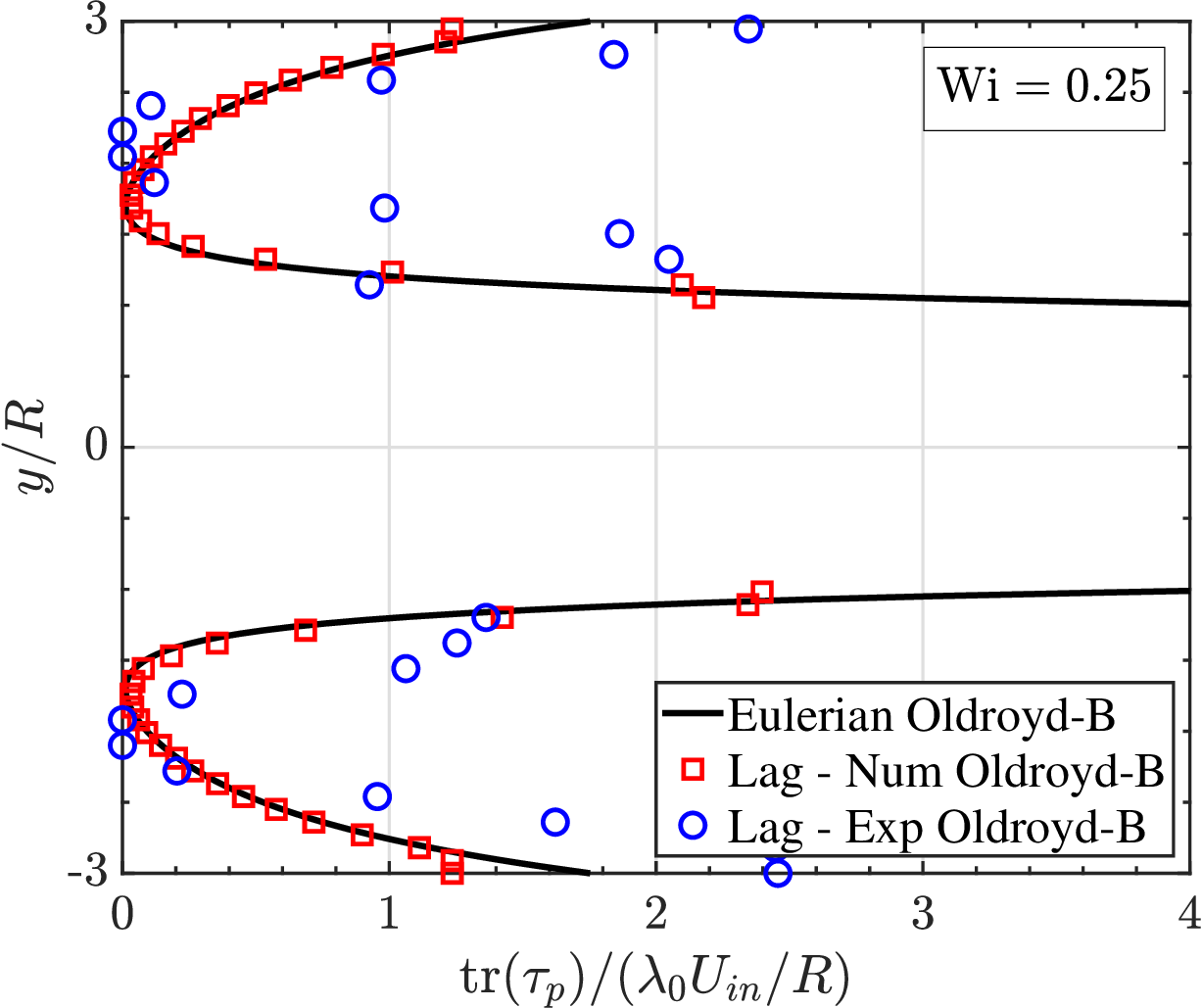}
        \caption{One cylinder}
        \label{fig:oldB_line_one_cylinder}
    \end{subfigure}
    \hfill
    \begin{subfigure}[t]{0.32\textwidth}
        \centering
        \includegraphics[width=\textwidth, trim={0cm 0cm 0cm 0cm}, clip]
        {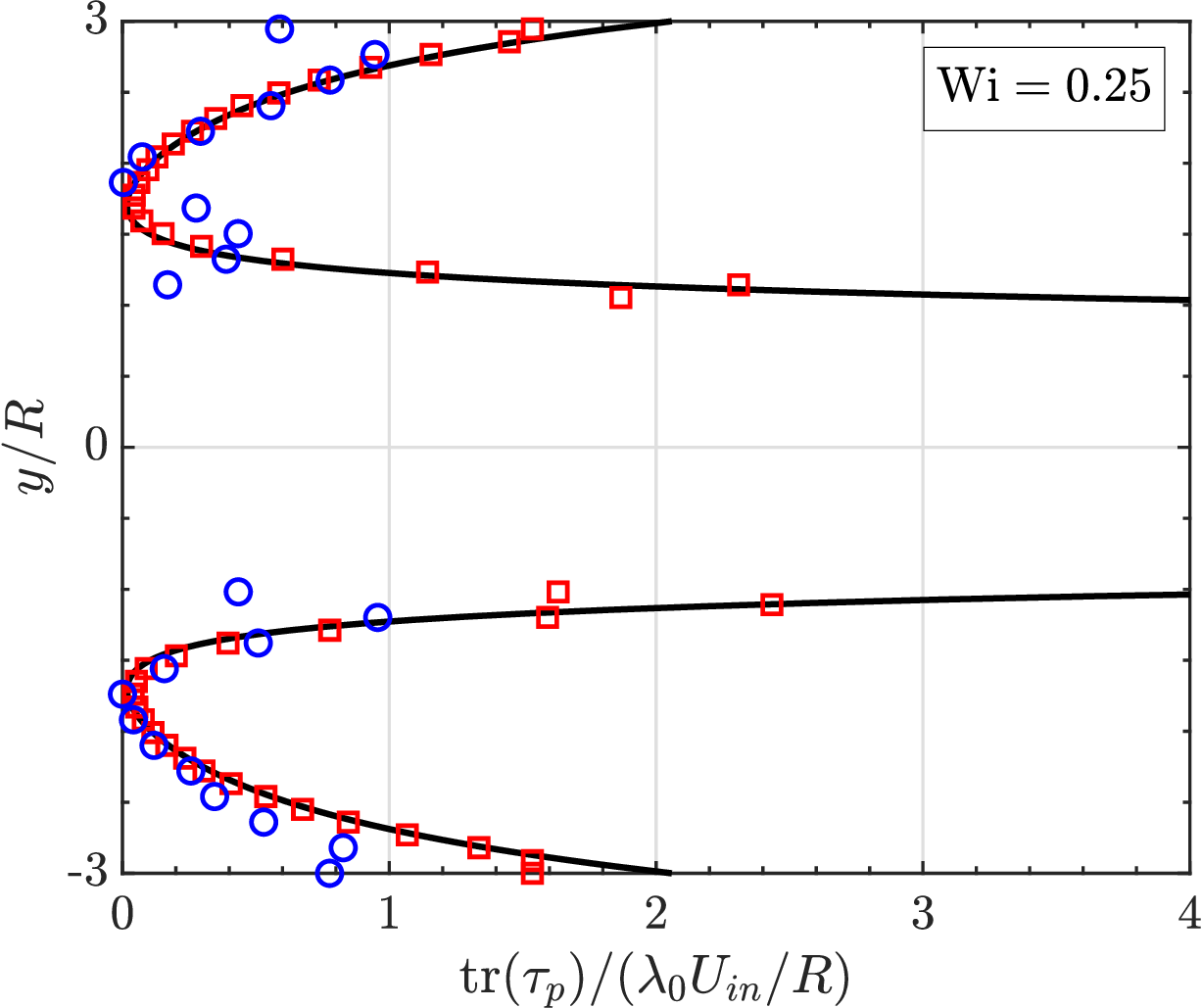}
        \caption{Two cylinders, left}
        \label{fig:oldB_line_LC}
    \end{subfigure}
    \hfill
    \begin{subfigure}[t]{0.32\textwidth}
        \centering
        \includegraphics[width=\textwidth, trim={0cm 0cm 0cm 0cm}, clip]
        {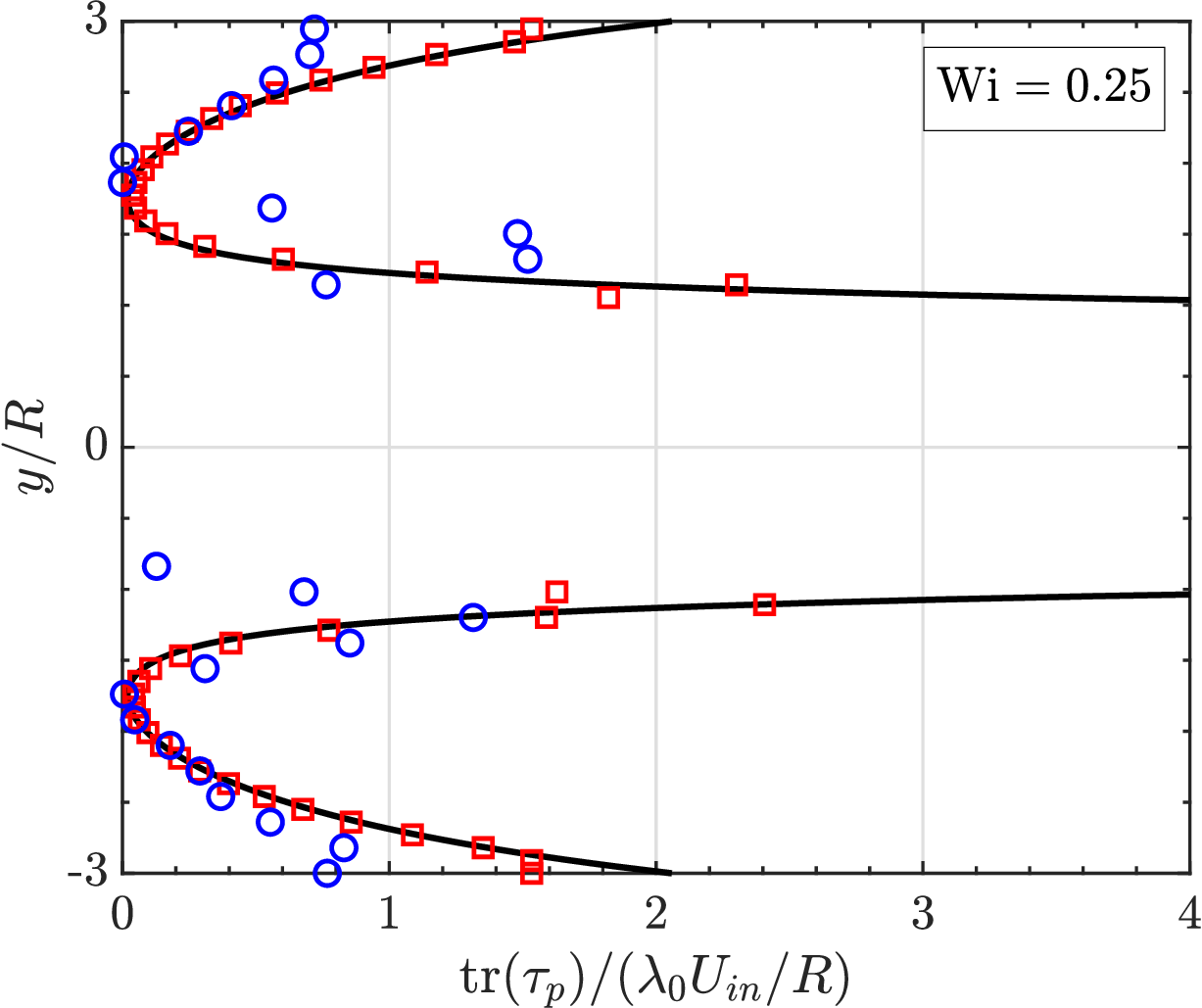}
        \caption{Two cylinders, right}
        \label{fig:oldB_line_RC}
    \end{subfigure}

    \vspace{0.15cm}

    % ===================== Row 2: circular profiles =====================
    \begin{subfigure}[t]{0.32\textwidth}
        \centering
        \includegraphics[width=\textwidth, trim={0cm 0cm 0cm 0cm}, clip]
        {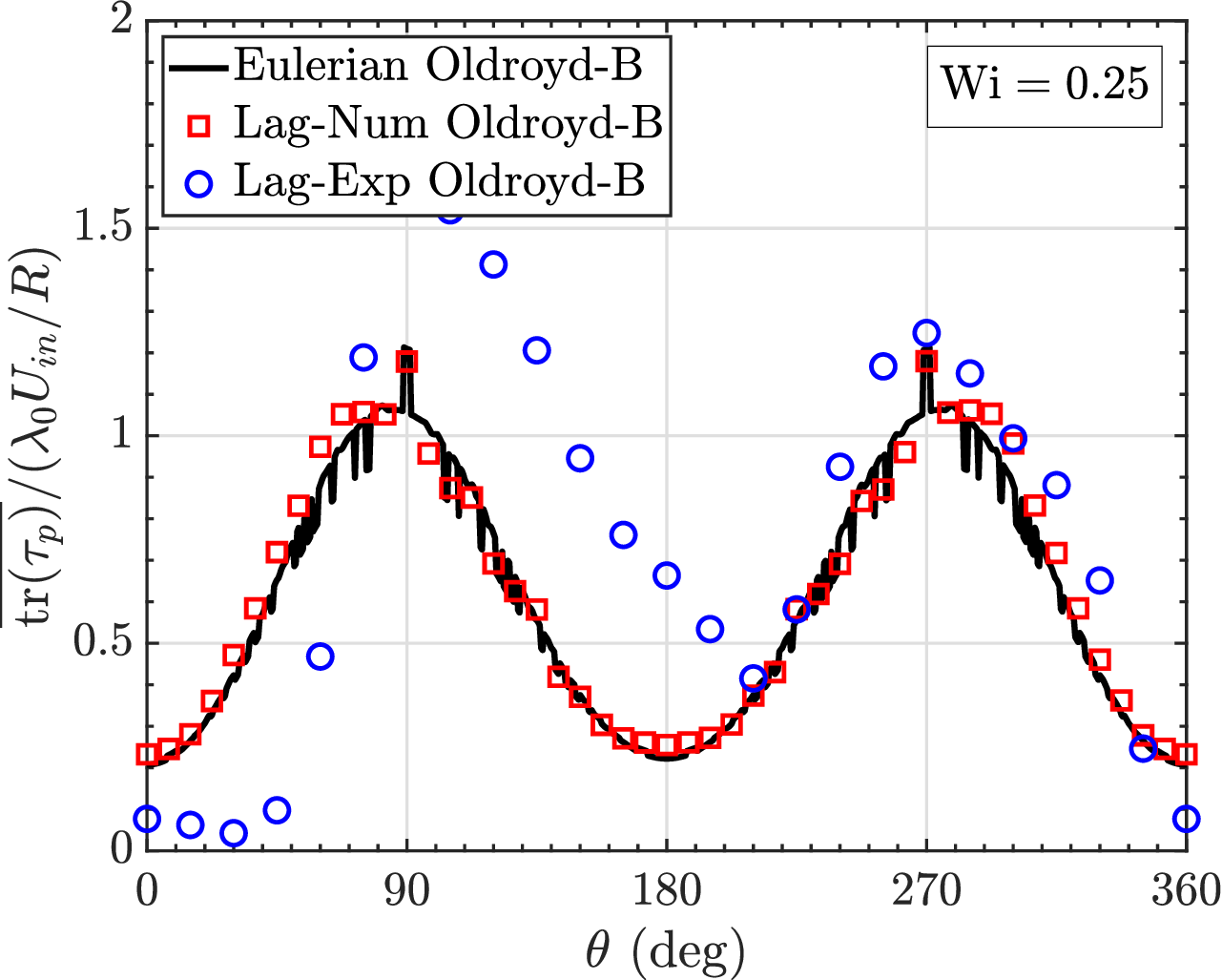}
        \caption{One cylinder}
        \label{fig:oldB_circle_one_cylinder}
    \end{subfigure}
    \hfill
    \begin{subfigure}[t]{0.32\textwidth}
        \centering
        \includegraphics[width=\textwidth, trim={0cm 0cm 0cm 0cm}, clip]
     {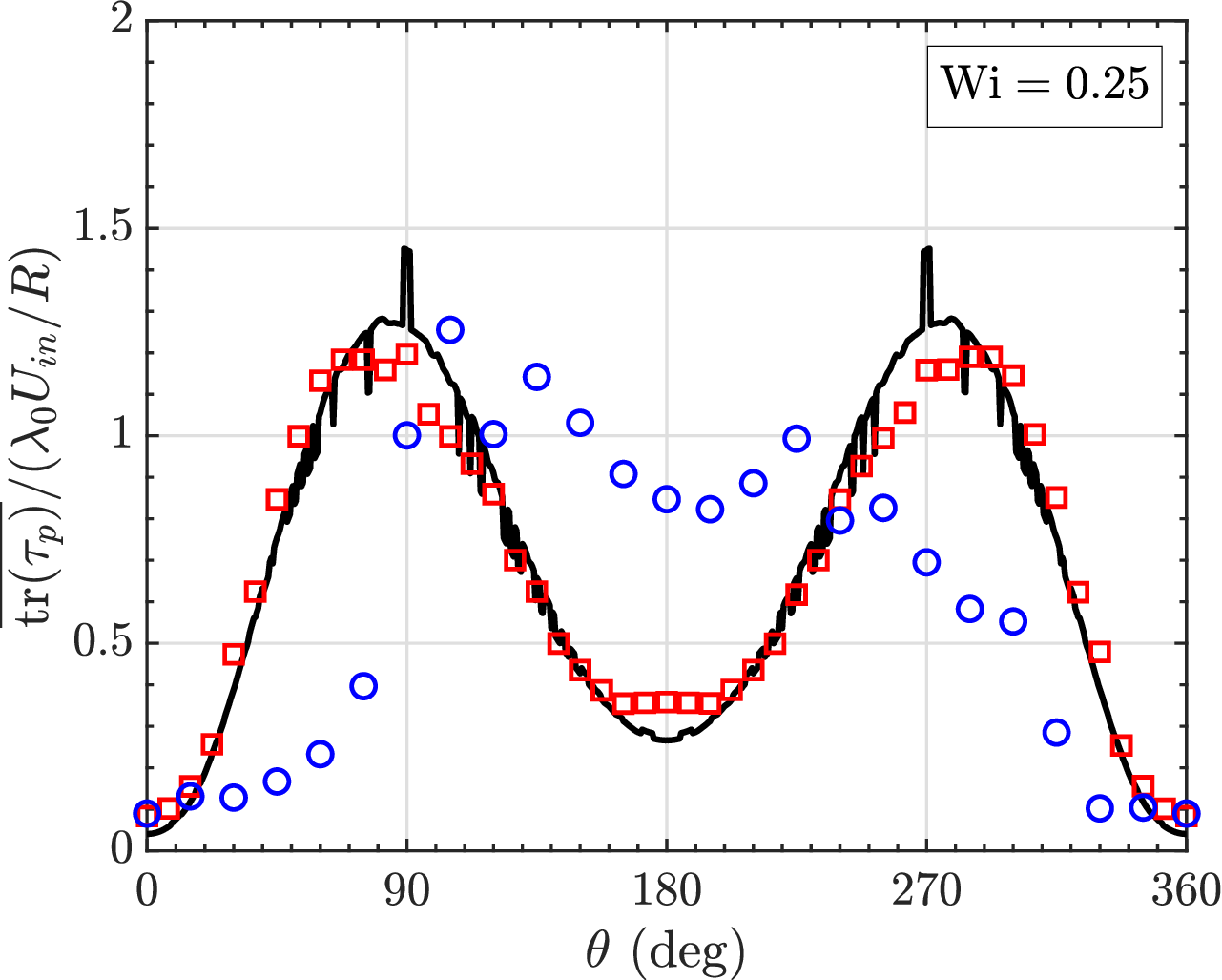}
        \caption{Two cylinders, left}
        \label{fig:oldB_circle_LC}
    \end{subfigure}
    \hfill
    \begin{subfigure}[t]{0.32\textwidth}
        \centering
        \includegraphics[width=\textwidth, trim={0cm 0cm 0cm 0cm}, clip]
     {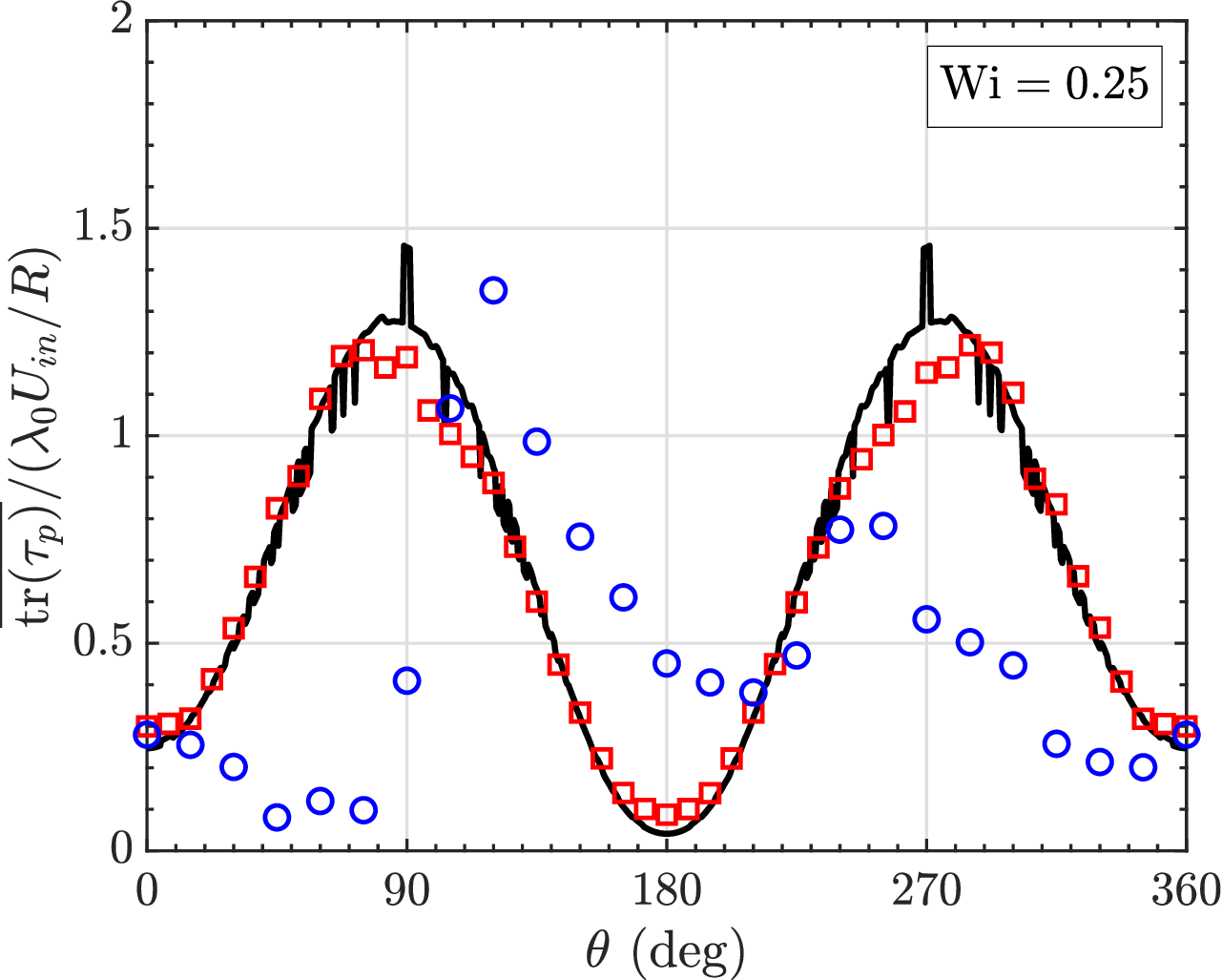}
        \caption{Two cylinders, right}
        \label{fig:oldB_circle_RC}
    \end{subfigure}
\caption{Profile comparison (see figure~\ref{fig:oldB_trTau_onevstwo})  of the dimensionless polymeric-stress trace for the Oldroyd-B model at $\mathrm{Wi}=0.25$. Red squares show the Lagrangian reconstruction using the numerical velocity field (Lag-Num), blue circles show the Lagrangian reconstruction using the experimental velocity field (Lag-Exp), and black lines show the Eulerian numerical reference solution. The top row shows $\mathrm{tr}(\boldsymbol{\tau}_p)$ profiles across $y/R$, while the bottom row shows radially-averaged azimuthal profiles around each cylinder over $r=40$--$60~\mu\mathrm{m}$. The first column corresponds to the one-cylinder case, and the second and third columns correspond to the left and right cylinders in the two-cylinder case. }
    \label{fig:oldB_profile_comparison}
\end{figure*}

To provide a more quantitative comparison, we examine the trace of the polymeric stress across the width of the channel and along an azimuthal profile that is radially averaged over the region $40\le r \le 60~\mu\mathrm{m}$ relative to the cylinder centers for both FENE-P and Oldroyd-B (figures~\ref{fig:fenep_profile_comparison} and~\ref{fig:oldB_profile_comparison} , respectively). This comparison illustrates the accuracy of the Lagrangian approach in predicting the stress in regions with high velocity gradients near solid surfaces. The Lagrangian approach captures the stress variation across the channel and within the annular region surrounding each cylinder with very good accuracy. The difference between the experimental result and the two numerical results for the radial-averaged profile is primarily due to low optical resolution near the cylinder surfaces.

 Overall, the stress fields reconstructed from the Lagrangian approach show good quantititative agreement with the Eulerian numerical results. This is observed for both the one-cylinder and two-cylinder geometries, and for both Oldroyd-B and FENE-P fluids. The comparison shows that the present formulation can recover the stress patterns in spatially non-uniform flows with mixed kinematics using either simulated or experimental velocity fields.

\begin{figure*}[!t]
    \centering
    \includegraphics[width=0.9\textwidth, trim={0cm 0cm 0cm 0cm}, clip]
    {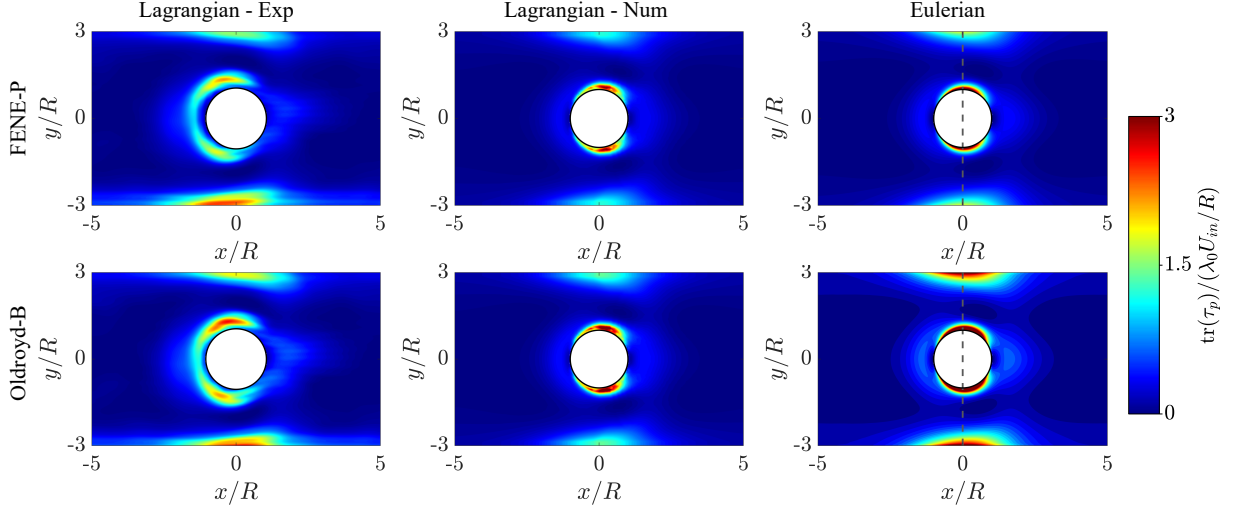}

    \caption{Comparison of the non-dimensional polymeric-stress trace for flow past one circular cylinder. The top row shows the FENE-P model at \(\mathrm{Wi}=0.68\), and the bottom row shows the Oldroyd-B model at the same $\mathrm{Wi}$ value. From left to right, the columns show the Lagrangian reconstruction using the experimental velocity field, the Lagrangian reconstruction using the numerical velocity field, and the Eulerian numerical reference solution. The dashed vertical lines indicate the
centreline locations for profiles of the polymeric-stress trace (see figure~\ref{fig:fene_oldb_profile_comparison}). }
\label{fig:one_cylinder_fenep_oldB_comparison}
\end{figure*}

\begin{figure*}[!t]
    \centering

    % ===================== Row 1: trTau vs y =====================
    \begin{subfigure}[t]{0.35\textwidth}
        \centering
        \includegraphics[width=\textwidth, trim={0cm 0cm 0cm 0cm}, clip]
        {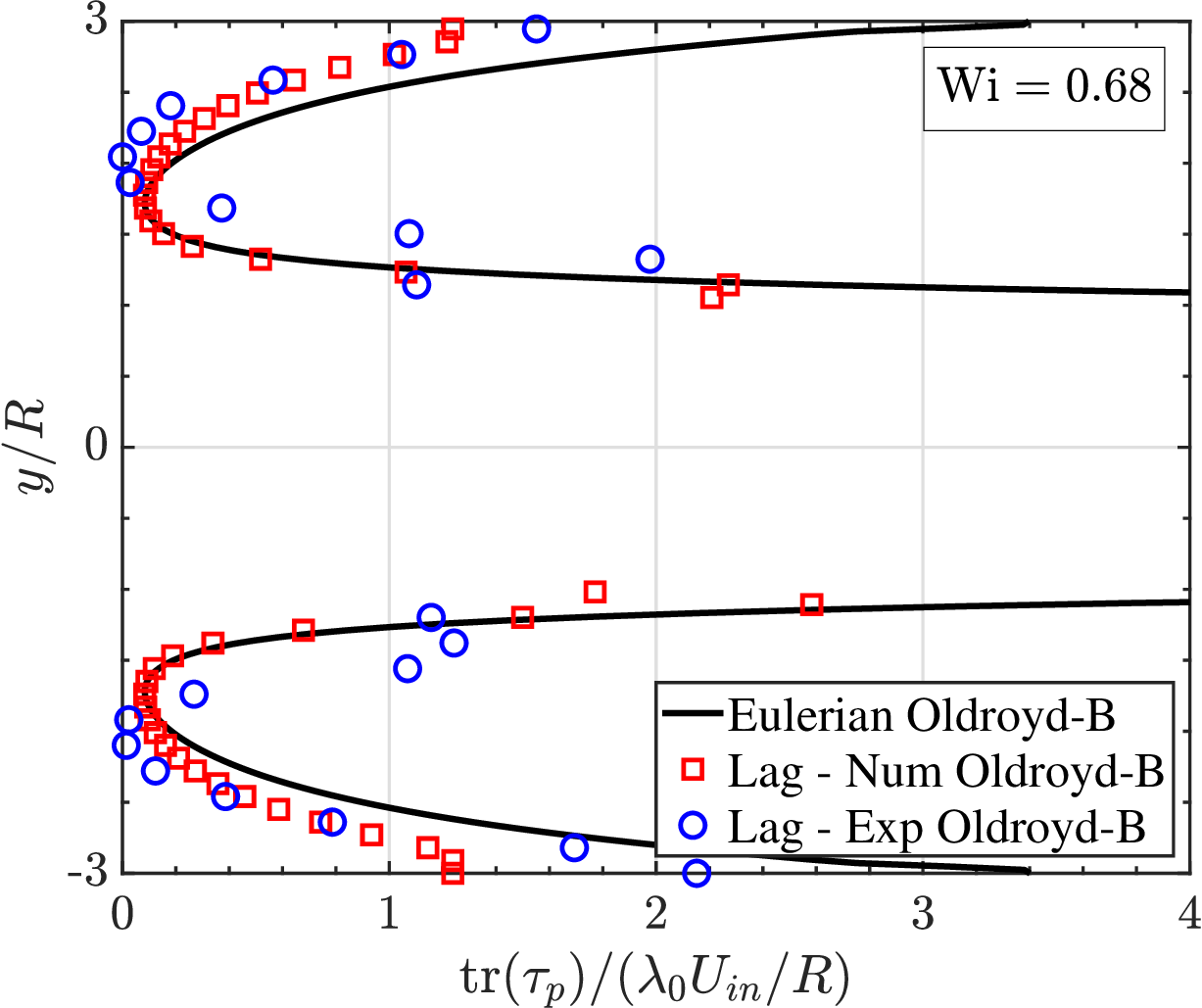}
        \caption{}
        \label{fig:oldB_line_one_cylinder}
    \end{subfigure}
    \begin{subfigure}[t]{0.35\textwidth}
        \centering
        \includegraphics[width=\textwidth, trim={0cm 0cm 0cm 0cm}, clip]
        {figures/line_onecylinder_fene.eps}
        \caption{}
        \label{fig:oldB_line_LC}
    \end{subfigure}
    
    % ===================== Row 2: circular profiles =====================
    \begin{subfigure}[t]{0.35\textwidth}
        \centering
        \includegraphics[width=\textwidth, trim={0cm 0cm 0cm 0cm}, clip]
        {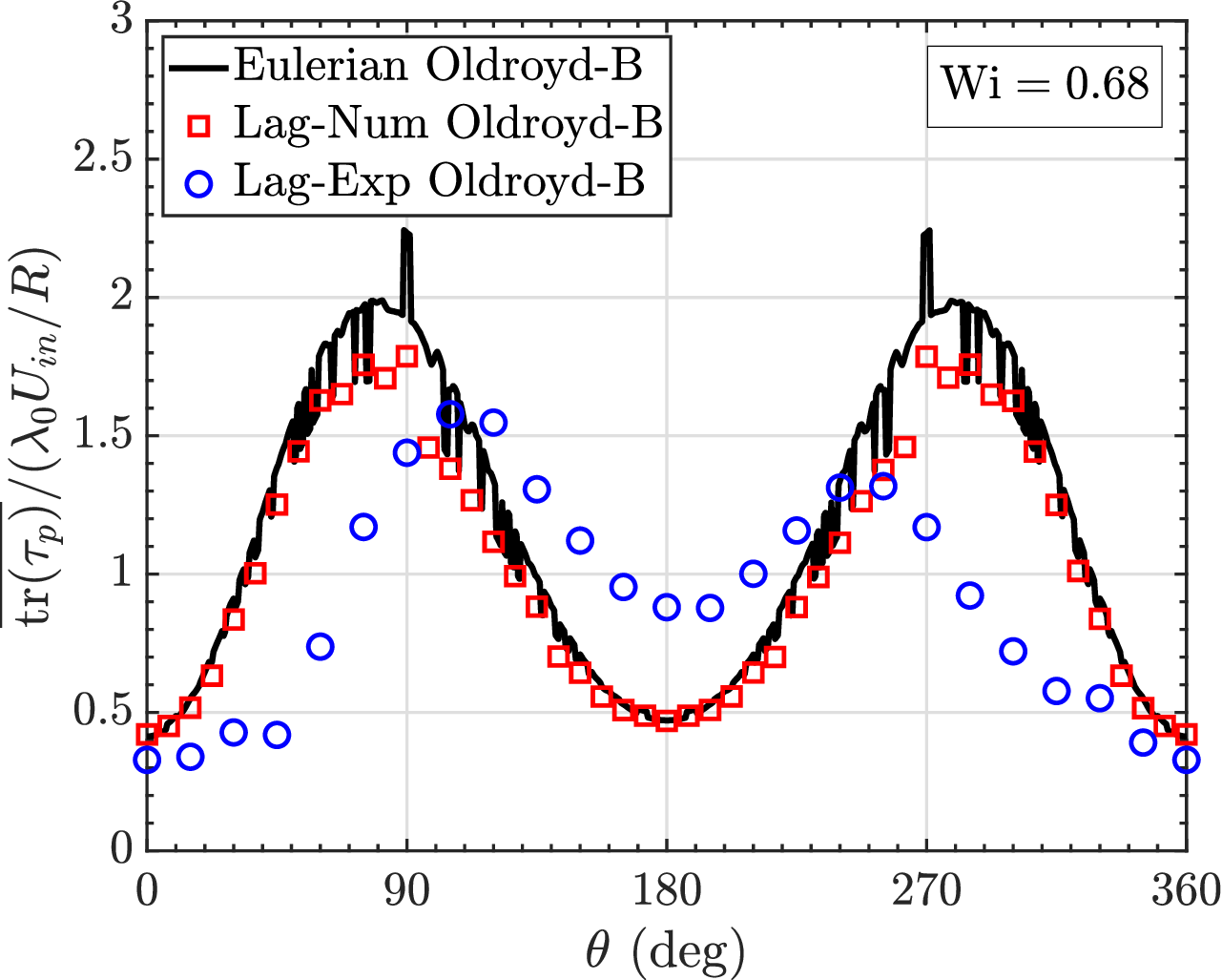}
        \caption{}
        \label{fig:oldB_circle_one_cylinder}
    \end{subfigure}
    \begin{subfigure}[t]{0.35\textwidth}
        \centering
        \includegraphics[width=\textwidth, trim={0cm 0cm 0cm 0cm}, clip]
     {figures/ave_onecylinder_fene.eps}
        \caption{}
        \label{fig:oldB_circle_LC}
    \end{subfigure}
    \caption{Profile comparison (see figure~\ref{fig:one_cylinder_fenep_oldB_comparison}) of the dimensionless polymeric-stress trace for the Oldroyd-B and FENE-P models at $\mathrm{Wi}=0.68$ for the one cylinder case. Red squares show the Lagrangian reconstruction using the numerical velocity field (Lag-Num), blue circles show the Lagrangian reconstruction using the experimental velocity field (Lag-Exp), and black lines show the Eulerian numerical reference solution. Panels (a) and (b) show $\mathrm{tr}(\boldsymbol{\tau}_p)$ profiles across $y/R$ for the Oldroyd-B and FENE-P models, respectively. Panels (c) and (d) show the corresponding radial-averaged azimuthal profiles around the cylinder over $r=40$--$60~\mu\mathrm{m}$.}
    \label{fig:fene_oldb_profile_comparison}
\end{figure*}

To further highlight the differences between the two constitutive models, the one cylinder case is also compared for the same (\(\mathrm{Wi}=0.68\); figure~\ref{fig:one_cylinder_fenep_oldB_comparison}). In this comparison, the same colorbar range is intentionally used for both models to make the difference in stress magnitude clearer. The Oldroyd-B model produces a slightly higher stress trace compared with the FENE-P model, especially near the surface of the cylindrical obstacle and in the top and bottom wall regions. This feature was also observed in the shear and planar flow problems above, and it is consistent with the unbounded Hookean spring response of the Oldroyd-B model. In contrast, the FENE-P model limits polymer stretching through finite extensibility \cite{bajaj2008coil}. These differences are highlighted quantitatively in the line and radially-averaged azimuthal profiles (figure~\ref{fig:fene_oldb_profile_comparison}). These results emphasize the importance of extending the Lagrangian formulation to FENE-P and other rheologies, where local deformation can lead to significant differences in polymeric stress.

 \FloatBarrier

 \section{Summary and conclusions}

In this work, a Lagrangian approach was formulated for determining the polymeric stress field in steady viscoelastic flows described by the FENE-P and Oldroyd-B constitutive equations. The formulation uses the deformation history of fluid elements to reconstruct the conformation tensor and polymeric stress from a known velocity field, established through experiments or simulations. 
A primary advantage of the present approach is that the stress field is determined without solving the Eulerian stress transport equation throughout the full domain and without comprehensive boundary conditions.

The Lagrangian formulation was first demonstrated for simple shear flow, for which analytical and asymptotic solutions are available. Excellent agreement was found between the Lagrangian calculation and the known FENE-P solutions. The  Oldroyd-B model diverged from the FENE-P model at large \(\mathrm{Wi}\) as expected, where the FENE-P trace is limited by finite extensibility relative to the continuous growth of the Oldroyd-B model. The method was also applied to planar channel flow having inhomogenous shear, where excellent agreement again was observed between the Lagrangian stress reconstruction and the Eulerian numerical solution.

The Lagrangian approach established here was further tested in channel flows with one and two circular cylindrical obstacles using both numerical and experimental velocity fields to assess its capability for such complex flow problems. As a measure of Lagrangian fluid deformation, the stretching fields were examined for the simulated and measured velocity fields, revealing similar deformation patterns. The results were consistent with prior works \cite{kumar_stress_2023,kumar2023lagrangian}, whereby high stretching regions around the cylinders were coincident with high polymeric stress regions. Importantly, the stretching field only reveals the approximate topology of the stress, but not the magnitude. The polymeric stress field was successfully determined through direct reconstruction using the new Lagrangian formulation.
For the obstacle flows, the Lagrangian approach, which uses the numerically determined velocity field, showed very good agreement with the Eulerian solution. The reconstruction based on the experimental velocity field also captured the essential stress patterns and their magnitude, although some differences were observed especially near surfaces. These discrepancies are partly attributed to finite resolution and measurement noise. However, constitutive models, including Oldroyd-B and FENE-P, also have limitations in modeling nuanced rheology of real fluids \cite{boyko2024perspective}.

Overall, the reconstructed stress fields accurately captured the main regions of high stress near the cylinder surfaces and in regions of strong local deformation. The comparison between FENE-P and Oldroyd-B also showed the importance of including finite extensibility, especially in flows with obstacles where strong mixed kinematics and extension lead to large polymeric stresses near solid surfaces. This is due to the fact that Oldroyd-B is a linear elastic model with unbounded polymer stretching. Hence, it can over-predict the stress magnitude in these high-deformation regions, while FENE-P provides a more physically limited response \cite{bajaj2008coil}.
The promising results presented in this study support the use of the Lagrangian formulation as a practical approach for reconstructing polymeric stress fields from a known velocity data. This is particularly useful for experimental flows, where velocity fields can be measured with relative ease but direct access to stress fields are are often challenging.

In the future, this method may be extended to more complex geometries and stronger flow conditions. With a diversity of advanced viscoelastic rheological models available \cite{boyko2024perspective},  determining whether such complex constitutive equations can be evaluated more conveniently in the Lagrangian frame could provide a significant advantage, without the need to solve fully coupled viscoelastic simulations in the Eulerian frame. Without loss of generality, the Lagrangian formulation presented here can be directly extended to unsteady flows, for which the velocity field varies with time. Since the present approach already follows the time history of fluid particles, it can also be applied to unsteady flow field data without changing the basic structure of the method. Taken together, the formulation presented here along with other recent works \cite{kumar_stress_2023,kumar2023lagrangian,stone2023note,snoeijer2020relationship} emphasize the importance and advantages of a Lagrangian perspective in analyzing and interpreting viscoelastic fluid flows.

\section*{Author contributions statement}

Conceptualization, M.M., J.S.G., and A.M.A.; computational methodology, M.M., R.G., J.S.G., and A.M.A.; experimental methodology, L.T. and M.T.; software, M.M.; validation, M.M., L.T., and M.T.; formal analysis, M.M., R.G., and L.T.; investigation, M.M., L.T., and M.T.; resources, J.S.G. and A.M.A.; data curation, M.M., L.T., and M.T.; writing---original draft preparation, M.M.; writing---review and editing, M.M., R.G., L.T., M.T., J.S.G., and A.M.A.; visualization, M.M.; supervision, J.S.G. and A.M.A.; project administration, J.S.G. and A.M.A. All authors have read and agreed to the published version of the manuscript.

\section*{Acknowledgments}
The work of M.M., A.A., and R.G was supported by .S. National Science Foundation (NSF)
through awards CBET-2341154 and CBET-2141349.
The work of J.S.G., M.T., and L.T. was supported by the NSF through awards CBET-2141349 and CMMI-2027410, and through the Gordon and Betty Moore Foundation, grant DOI 10.37807/GBMF13806.

\newpage

\appendix

\section*{Appendix}
\addcontentsline{toc}{section}{Appendix}

\subsection{Component-wise Form of the Lagrangian Conformation Tensor}
\label{appendix:componentwise_conformation}

For the two-dimensional case, let
\[
\mathbf{A}_L(t)=
\begin{bmatrix}
A_{11}(t) & A_{12}(t) \\
A_{21}(t) & A_{22}(t)
\end{bmatrix},
\qquad
\mathbf{F}(t)=
\begin{bmatrix}
F_{11}(t) & F_{12}(t) \\
F_{21}(t) & F_{22}(t)
\end{bmatrix}.
\]
The component form of equation~(\ref{7e}) is then given by
\begin{multline}
\begin{bmatrix}
A_{11}(t) & A_{12}(t) \\
A_{21}(t) & A_{22}(t)
\end{bmatrix}
=
\begin{bmatrix}
F_{11}(t) & F_{21}(t) \\
F_{12}(t) & F_{22}(t)
\end{bmatrix}
\begin{bmatrix}
1 & 0 \\
0 & 1
\end{bmatrix}
\begin{bmatrix}
F_{11}(t) & F_{12}(t) \\
F_{21}(t) & F_{22}(t)
\end{bmatrix}
\exp\!\left(
-\int_0^t
\frac{1}{\lambda_0}
f\!\left(\operatorname{tr}(\mathbf{A}_L(s))\right)\,ds
\right)
\\
+
\frac{a}{\lambda_0}
\begin{bmatrix}
F_{11}(t) & F_{21}(t) \\
F_{12}(t) & F_{22}(t)
\end{bmatrix}
\exp\!\left(
-\int_0^t
\frac{1}{\lambda_0}
f\!\left(\operatorname{tr}(\mathbf{A}_L(s))\right)\,ds
\right)
\\
\times
\left[
\int_0^t
\exp\!\left(
\int_0^{t'}
\frac{1}{\lambda_0}
f\!\left(\operatorname{tr}(\mathbf{A}_L(s))\right)\,ds
\right)
\frac{1}{\left(F_{11}(t')F_{22}(t')-F_{12}(t')F_{21}(t')\right)^2}
\right.
\\
\left.
\times
\begin{bmatrix}
F_{22}(t') & -F_{21}(t') \\
-F_{12}(t') & F_{11}(t')
\end{bmatrix}
\begin{bmatrix}
F_{22}(t') & -F_{12}(t') \\
-F_{21}(t') & F_{11}(t')
\end{bmatrix}
\,dt'
\right]
\begin{bmatrix}
F_{11}(t) & F_{12}(t) \\
F_{21}(t) & F_{22}(t)
\end{bmatrix}.
\end{multline}
After carrying out the matrix multiplications, we have
\begin{multline}
\begin{bmatrix}
A_{11}(t) & A_{12}(t) \\
A_{21}(t) & A_{22}(t)
\end{bmatrix}
=
\begin{bmatrix}
F_{11}^2(t)+F_{21}^2(t) &
F_{11}(t)F_{12}(t)+F_{21}(t)F_{22}(t) \\
F_{11}(t)F_{12}(t)+F_{21}(t)F_{22}(t) &
F_{12}^2(t)+F_{22}^2(t)
\end{bmatrix}
\exp\!\left(
-\int_0^t
\frac{1}{\lambda_0}
f\!\left(\operatorname{tr}(\mathbf{A}_L(s))\right)\,ds
\right)
\\
+
\frac{a}{\lambda_0}
\begin{bmatrix}
F_{11}(t) & F_{21}(t) \\
F_{12}(t) & F_{22}(t)
\end{bmatrix}
\exp\!\left(
-\int_0^t
\frac{1}{\lambda_0}
f\!\left(\operatorname{tr}(\mathbf{A}_L(s))\right)\,ds
\right)
\\
\times
\left[
\int_0^t
\exp\!\left(
\int_0^{t'}
\frac{1}{\lambda_0}
f\!\left(\operatorname{tr}(\mathbf{A}_L(s))\right)\,ds
\right)
\frac{1}{
\left(F_{11}(t')F_{22}(t')-F_{12}(t')F_{21}(t')\right)^2
}
\right.
\\
\left.
\times
\begin{bmatrix}
F_{22}^2(t')+F_{21}^2(t') &
-F_{22}(t')F_{12}(t')-F_{21}(t')F_{11}(t') \\
-F_{22}(t')F_{12}(t')-F_{21}(t')F_{11}(t') &
F_{12}^2(t')+F_{11}^2(t')
\end{bmatrix}
\,dt'
\right]
\begin{bmatrix}
F_{11}(t) & F_{12}(t) \\
F_{21}(t) & F_{22}(t)
\end{bmatrix}.
\end{multline}
Using the memory factor \(\mathcal{M}(t)\) defined in the main text, the same expression can be written more compactly as
\begin{multline}
\begin{bmatrix}
A_{11}(t) & A_{12}(t) \\
A_{21}(t) & A_{22}(t)
\end{bmatrix}
=
\begin{bmatrix}
F_{11}^2(t)+F_{21}^2(t) &
F_{11}(t)F_{12}(t)+F_{21}(t)F_{22}(t) \\
F_{11}(t)F_{12}(t)+F_{21}(t)F_{22}(t) &
F_{12}^2(t)+F_{22}^2(t)
\end{bmatrix}
\mathcal{M}(t)
\\
+
\frac{a}{\lambda_0}\mathcal{M}(t)
\begin{bmatrix}
F_{11}(t) & F_{21}(t) \\
F_{12}(t) & F_{22}(t)
\end{bmatrix}
\left[
\int_0^t
\frac{1}{\mathcal{M}(t')}
\mathbf{G}(t')\,dt'
\right]
\begin{bmatrix}
F_{11}(t) & F_{12}(t) \\
F_{21}(t) & F_{22}(t)
\end{bmatrix},
\end{multline}
where
\begin{equation}
\mathbf{G}(t')
=
\frac{1}
{\left(F_{11}(t')F_{22}(t')-F_{12}(t')F_{21}(t')\right)^2}
\begin{bmatrix}
F_{22}^2(t')+F_{21}^2(t') &
-F_{22}(t')F_{12}(t')-F_{21}(t')F_{11}(t') \\
-F_{22}(t')F_{12}(t')-F_{21}(t')F_{11}(t') &
F_{12}^2(t')+F_{11}^2(t')
\end{bmatrix}.
\end{equation}
Using the compact notation introduced in the main text, the diagonal components can be written from the expanded form as
\begin{multline}
A_{11}(t)
=
\left(F_{11}^2(t)+F_{21}^2(t)\right)\mathcal{M}(t)
\\
+\frac{a}{\lambda_0}\mathcal{M}(t)
\int_{0}^{t}
\frac{1}{\mathcal{M}(t')}
\bigg[
F_{11}(t)
\left(
\frac{
\left(F_{22}^2(t')+F_{21}^2(t')\right)F_{11}(t)
+\left(-F_{22}(t')F_{12}(t')-F_{21}(t')F_{11}(t')\right)F_{21}(t)
}{
\left(F_{11}(t')F_{22}(t')-F_{12}(t')F_{21}(t')\right)^2
}
\right)
\\
+
F_{21}(t)
\left(
\frac{
\left(F_{12}^2(t')+F_{11}^2(t')\right)F_{21}(t)
+\left(-F_{22}(t')F_{12}(t')-F_{21}(t')F_{11}(t')\right)F_{11}(t)
}{
\left(F_{11}(t')F_{22}(t')-F_{12}(t')F_{21}(t')\right)^2
}
\right)
\bigg]\,dt' .
\end{multline}

\begin{multline}
A_{22}(t)
=
\left(F_{12}^2(t)+F_{22}^2(t)\right)\mathcal{M}(t)
\\
+\frac{a}{\lambda_0}\mathcal{M}(t)
\int_{0}^{t}
\frac{1}{\mathcal{M}(t')}
\bigg[
F_{12}(t)
\left(
\frac{
\left(F_{22}^2(t')+F_{21}^2(t')\right)F_{12}(t)
+\left(-F_{22}(t')F_{12}(t')-F_{21}(t')F_{11}(t')\right)F_{22}(t)
}{
\left(F_{11}(t')F_{22}(t')-F_{12}(t')F_{21}(t')\right)^2
}
\right)
\\
+
F_{22}(t)
\left(
\frac{
\left(F_{12}^2(t')+F_{11}^2(t')\right)F_{22}(t)
+\left(-F_{22}(t')F_{12}(t')-F_{21}(t')F_{11}(t')\right)F_{12}(t)
}{
\left(F_{11}(t')F_{22}(t')-F_{12}(t')F_{21}(t')\right)^2
}
\right)
\bigg]\,dt' .
\end{multline}
The diagonal components \(A_{11}(t)\) and \(A_{22}(t)\) obtained above are the quantities used in the nonlinear trace equation. Since the trace appears inside the memory factor, these expressions are evaluated for a trial value of \(\operatorname{tr}(\mathbf{A}_L)\), and the scalar equation for the trace is solved iteratively using Brent's method, as described in the algorithm section. The same formulation can also be extended to a \(3\times3\) tensor by taking
\[
\mathbf{A}_L(t)=
\begin{bmatrix}
    A_{11}(t) & A_{12}(t) & A_{13}(t) \\
    A_{21}(t) & A_{22}(t) & A_{23}(t) \\
    A_{31}(t) & A_{32}(t) & A_{33}(t)
\end{bmatrix},
\qquad
\mathbf{F}(t)=
\begin{bmatrix}
    F_{11}(t) & F_{12}(t) & F_{13}(t) \\
    F_{21}(t) & F_{22}(t) & F_{23}(t) \\
    F_{31}(t) & F_{32}(t) & F_{33}(t)
\end{bmatrix}.
\]

\subsection{Computational Cost of Lagrangian Stress Reconstruction}
\label{computational_cost}

The computational cost of the Lagrangian approach was evaluated for the one-cylinder case. The reported times correspond only to reconstruction of the polymeric stress field from a prescribed numerical velocity field; the time required to obtain the velocity field is not included. All calculations were performed on Purdue University's Negishi high-performance computing cluster using AMD EPYC 7763 (``Milan'') processors.

The results reported in Table~\ref{tab:lagrangian_efficiency} correspond to the FENE-P model, for which the nonlinear scalar trace equation was solved iteratively using Brent's method, as described in Section~\ref{iteratviesolverfene}. The MATLAB implementation used 4, 8, 16, and 32 parallel workers. Given the velocity field, the Lagrangian method reconstructed the FENE-P polymeric stress field in as little as 58.15~s using 32 workers. The corresponding results for the two-cylinder case were similar and are not repeated here for brevity.

For context, the corresponding coupled Eulerian simulation in RheoTool required 17318~s to solve the flow field and polymeric stress field simultaneously. This time includes the coupled solution of the velocity, pressure, and constitutive equations and is therefore not presented as a direct runtime comparison with the Lagrangian reconstruction. Rather, the results in Table~\ref{tab:lagrangian_efficiency} demonstrate that, once a velocity field is available, the Lagrangian approach can efficiently reconstruct the polymeric stress field.

For the Oldroyd-B model, the Lagrangian formulation is explicit and does not require an iterative solution of the trace equation. Consequently, its computational cost is substantially lower than that of the iterative FENE-P calculation. In the present parallel MATLAB calculations, the Oldroyd-B stress reconstruction required less than 10~s for all worker counts considered. Therefore, the Oldroyd-B results are not included in Table~\ref{tab:lagrangian_efficiency}.

 \begin{table}[H]
\centering
\caption{Wall-clock runtime of the Lagrangian stress reconstruction from a prescribed velocity field.}
\label{tab:lagrangian_efficiency}

\small
\renewcommand{\arraystretch}{1.15}
\setlength{\tabcolsep}{6pt}

\begin{tabular}{lrrr}
\hline
Approach & Cores & Time (s) & Speedup \\
\hline
Lagrangian & 4  & 231.30 & 74.87 \\
 & 8  & 131.01 & 132.19 \\
 & 16 & 79.25  & 218.52 \\
 & 32 & 58.15  & 297.82 \\
\hline
\end{tabular}

\end{table}

\newpage

\bibliography{bibliography}
\bibliographystyle{ieeetr}

\end{document}